\documentclass[final]{IEEEtran}
\usepackage{cite}
\usepackage{graphicx}
\usepackage{picinpar}
\usepackage[cmex10]{amsmath}
\usepackage{amsmath,amsfonts,amssymb}
\usepackage{subfigure}

\usepackage{algorithm}
\usepackage{algorithmic}
\usepackage{stfloats}
\usepackage{bm}

\usepackage{stfloats}
\usepackage{picins}

\begin{document}
\newtheorem{lemma}{Lemma}
\newtheorem{corol}{Corollary}
\newtheorem{theorem}{Theorem}
\newtheorem{proposition}{Proposition}
\newtheorem{definition}{Definition}
\newcommand{\e}{\begin{equation}}
\newcommand{\ee}{\end{equation}}
\newcommand{\eqn}{\begin{eqnarray}}
\newcommand{\eeqn}{\end{eqnarray}}

\renewcommand{\algorithmicrequire}{ \textbf{Input:}} 
\renewcommand{\algorithmicensure}{ \textbf{Output:}} 

\title{2D Unitary ESPRIT Based Super-Resolution Channel Estimation for Millimeter-Wave Massive MIMO with Hybrid Precoding}

\author{Anwen Liao, Zhen Gao,~\IEEEmembership{Member,~IEEE}, Yongpeng Wu,~\IEEEmembership{Senior Member,~IEEE}, Hua Wang,~\IEEEmembership{Member,~IEEE}, \\
and Mohamed-Slim Alouini,~\IEEEmembership{Fellow,~IEEE}
\vspace*{-3.0mm}
\thanks{A. Liao and Z. Gao are with both Advanced Research Institute of Multidisciplinary Science (ARIMS) and School of Information and Electronics,
Beijing Institute of Technology (BIT), Beijing 100081, China (E-mail: \{2220160356, gaozhen16\}@bit.edu.cn).}
\thanks{Y. Wu is with the Department of Electronic Engineering,
Shanghai Jiao Tong University, Shanghai 200240, China (E-mail: yongpeng.wu2016@gmail.com).}
\thanks{H. Wang is with Research Institute of Telecommunication Technology, School of Information and Electronics,
Beijing Institute of Technology (BIT), Beijing 100081, China (E-mails: wanghua@bit.edu.cn).}
\thanks{M. -S. Alouini is with the Electrical Engineering Program, Division of Physical Sciences and Engineering,
King Abdullah University of Science and Technology (KAUST), Thuwal, Makkah Province, Saudi Arabia (E-mail: slim.alouini@kaust.edu.sa).}
\thanks{This work was supported by the National Natural Science Foundation
of China (Grant Nos. 61701027 and 61671294). (Corresponding author: Zhen Gao)} 
}

\maketitle

\begin{abstract}
Millimeter-wave (mmWave) massive multiple-input multiple-output (MIMO) with hybrid precoding is a promising technique for the future 5G wireless communications.
Due to a large number of antennas but a much smaller number of radio frequency (RF) chains, estimating the high-dimensional mmWave massive MIMO channel will bring the large pilot overhead.
To overcome this challenge, this paper proposes a super-resolution channel estimation scheme based on two-dimensional (2D) unitary ESPRIT algorithm.
By exploiting the angular sparsity of mmWave channels, the continuously distributed angle of arrivals/departures (AoAs/AoDs) can be jointly estimated with high accuracy.
Specifically, by designing the uplink training signals at both base station (BS) and mobile station (MS), we first use low pilot overhead to estimate a low-dimensional effective channel, which has the same shift-invariance
of array response as the high-dimensional mmWave MIMO channel to be estimated.
From the low-dimensional effective channel, the super-resolution estimates of AoAs and AoDs can be jointly obtained by exploiting the 2D unitary ESPRIT channel estimation algorithm.
Furthermore, the associated path gains can be acquired based on the least squares (LS) criterion.
Finally, we can reconstruct the high-dimensional mmWave MIMO channel according to the obtained AoAs, AoDs, and path gains.
Simulation results have confirmed that the proposed scheme is superior to conventional schemes with a much lower pilot overhead. 
\end{abstract}
\vspace*{-1mm}
\vspace*{-2.0mm}

\IEEEpeerreviewmaketitle

\section{Introduction}
Millimeter-wave (mmWave) massive multiple-input multiple-output (MIMO) has been widely regarded as one of the most important technologies for the next generation wireless communications due to its large under-utilized bandwidth at mmWave frequency band \cite{{overview:Heath}}.
For mmWave MIMO systems, a large number of antennas with the small form factor can be deployed at the base station (BS) and mobile station (MS) for achieving the large array gain to combat the severe path loss of mmWave channels.
However, for the large antenna array in mmWave MIMO, the conventional full digital precoding can be unaffordable, since each antenna requires the associated expensive radio frequency (RF) chain and high power-consumed analog-to-digital converters (ADCs) \cite{{overview:Heath},{Energy:efficientGao}}.
To reduce the hardware cost and power consumption as well as achieve the spatial multiplexing, the hybrid precoding with a much smaller number of RF chains than that of antennas has been proposed.
The hybrid precoding consists of the digital precoding at baseband and analog precoding at RF front end, which can be exploited to realize both beamforming and spatial multiplexing \cite{{MIMO:PCS},{Hybrid:MIMO}}.

For mmWave MIMO, the accurate channel state information (CSI) acquisition is essential to realize the sophisticated signal processing.
However, the high-dimensional channel coefficients to be estimated but the low-dimensional effective observations can be used will impose the prohibitively high pilot overhead on channel estimation. To this end, there have been several channel estimation schemes proposed for mmWave massive MIMO \cite{{Auxiliary},{Beamspace},{LimitedRF},{AdaptiveCS},{ACS},{OMP},{Subspace},{LSE},{Overlapped}}.
Specifically, based on the least squares (LS) estimation and sparse message passing, the authors in \cite{{LSE}} have proposed a least square estimation and sparse message passing-based channel estimation algorithm, which can iteratively detect the location and the value of non-zero entries of sparse channel vector.
In \cite{{Overlapped}}, based on the design of overlapped beam pattern, a fast channel estimation algorithm was first proposed to increase the amount of information and reduce the required channel estimation time, and then a rate-adaptive channel estimation algorithm was further proposed for the high probability of estimation error.
Unfortunately, the schemes proposed in both \cite{{LSE}} and \cite{{Overlapped}} are limited to the the conventional full digital precoding.
%
%
In \cite{{Beamspace}}, a support detection based channel estimation scheme was developed to detect the non-zero entries of sparse mmWave channels by utilizing the low rank and spatial correlation characteristics of mmWave beamspace channel.
In \cite{{LimitedRF}}, a two-way training scheme was proposed to achieve energy-based antenna selections at both BS and MS.
However, both \cite{{Beamspace}} and \cite{{LimitedRF}} only consider the lens antenna arrays based mmWave MIMO systems, and they may not be directly used for the hybrid precoding systems.

For mmWave MIMO with hybrid precoding, the authors in \cite{{Auxiliary}} have designed the auxiliary beam pairs to estimate the angle of arrivals/departures (AoAs/AoDs) of the channels, which can be used for the directional initial access process and facilitate data channel spatial multiplexing in the hybrid precoding.
In \cite{{Subspace}}, the main singular values and singular subspaces of the high-dimensional mmWave MIMO channels can be directly acquired by exploiting the Arnoldi iteration of Krylov subspace method, while multiple echoing operations required between the BS and MS can introduce much noise and degrade the estimation performance.
By exploiting the angular sparsity of mmWave channels, the compressed sensing (CS)-based schemes have been proposed to reduce pilot overhead \cite{{AdaptiveCS},{ACS},{OMP}}.
Specifically, the authors in \cite{{ACS}} proposed an adaptive CS (ACS)-based channel estimation scheme, where a hierarchical multi-resolution beamforming codebook is designed.
In \cite{{OMP}}, an orthogonal matching pursuit (OMP)-based scheme was proposed, where a nonuniformly quantized grid in the angle domain is adopted to reduce the correlation of redundant dictionary matrix and improve the performance of channel estimation.
On the other hand, most of existing CS-based channel estimation schemes assume the discrete AoAs/AoDs, which will bring a certain performance loss when estimating the practical continuously distributed AoAs/AoDs.

To estimate the practical AoAs/AoDs with the continuous other than discrete values, the classical spatial spectrum estimation algorithms, such as estimating signal parameters via rotational invariance techniques (ESPRIT) algorithms \cite{{esprit},{2Du:esprit}}, can be used.
These algorithms can facilitate the acquisition of the super-resolution estimates of the AoAs and AoDs with high accuracy.
The conventional ESPRIT algorithms require the observation samples to preserve the shift-invariance of array response, and they are widely used in the full digital array, where each baseband observation is directly sampled from the signal received by one antenna after the associated processing.
However, for mmWave MIMO with hybrid precoding, each baseband observation mixes the signals from different antennas via the RF phase shift network.
This indicates that directly using the conventional ESPRIT algorithms can be challenging, since the shift-invariance of array response is destroyed by the RF phase shift network.

With this in mind, this paper first designs the training signals at both BS and MS for channel estimation, so that we can obtain a low-dimensional effective channel having the shift-invariance of array response with low pilot overhead.
Then, we obtain the super-resolution estimates of the AoAs and AoDs jointly by exploiting the shift invariance of array response preserved in this low-dimensional effective channel matrix with the aid of 2D unitary ESPRIT based channel estimation algorithm.
Moreover, the path gains are estimated by applying the LS estimator.
Finally, the high-dimensional mmWave MIMO channel is reconstructed according to the acquired AoAs, AoDs, and the corresponding path gains.
Simulation results have shown that the proposed 2D unitary ESPRIT based super-resolution channel estimation scheme achieves better performance than conventional schemes with a reduced pilot overhead.
To the best of our knowledge, this is the first work to directly achieve the super-resolution estimates of AoAs and AoDs pairs by exploiting the classical spectral estimation algorithms, including ESPRIT-type and multiple signal classification (MUSIC)-type algorithms, for mmWave massive MIMO systems with hybrid precoding.

\textit{Notations}: The italic boldface lower and upper-case symbols denote column vectors and matrices, respectively.
Superscripts $(\cdot)^*$, $(\cdot)^T$, $(\cdot)^H$, $(\cdot)^{-1}$, $(\cdot)^\dagger$ denote the conjugate, transpose, conjugate transpose, inversion, and Moore-Penrose inversion operators, respectively.
${\left\| {\bm{a}} \right\|_2}$ and ${\left\| {\bm{A}} \right\|_F}$ are the ${l_2}$ norm of ${\bm{a}}$ and the Frobenius norm of ${\bm{A}}$, respectively.
$\otimes$ and $\odot$ denote the Kronecker product and Khatri-Rao product (also named the column wise Kronecker product), respectively.
${{\rm{diag}}( {\bm{a}} )}$ is a diagonal matrix with elements of ${\bm{a}}$ on its diagonal and ${{\rm{diag}}( {\bm{A}} )}$ denotes a block diagonal matrix with subblocks of ${\bm{A}}$ on its block diagonal if ${\bm{A}}$ is made up of multiple subblocks.
${{\rm{vec}}( {\bm{A}} )}$ denotes the vectorization operation of ${\bm{A}}$.
${{\bm{I}}_n} \ ( {{{\bm{O}}_{m \times n}}} )$  denotes an identity (null) matrix with size $n \times n \ ( {m \times n} )$, and ${{\bm{U}}_n}$ denotes a unitary matrix with size $n \times n$.
$\rm{Re\{\cdot\}}$ and $\rm{Im\{\cdot\}}$ denote the real part and imaginary part of the corresponding arguments, respectively.
${{\bm{J}}_n}$ is an exchange matrix with size $n \times n$ that reverses the order of rows of ${{\bm{I}}_n}$.
The expectation and determinant operators are denoted by $\mathbb{E}(\cdot)$ and ${{\rm{det}}(\cdot)}$, respectively.

\section{System Model}

We consider a typical mmWave massive MIMO uplink system with the hybrid precoding, where both MS and BS are equipped with $N_{\rm{MS}}$ and $N_{\rm{BS}}$ antennas but only $N_{\rm{RF}}^{\rm{MS}}$ and $N_{\rm{RF}}^{\rm{BS}}$ RF chains, respectively \cite{{overview:Heath},{MIMO:PCS},{Hybrid:MIMO},{Auxiliary},{AdaptiveCS},{ACS},{OMP},{Subspace}}.
$N_{\rm{S}}$ independent data streams are employed by MS and BS, such that ${N_{\rm{S}}} \le N_{\rm{RF}}^{\rm{MS}} \le {N_{\rm{MS}}}$ and $N_{\rm{S}} \le N_{\rm{RF}}^{\rm{BS}} \le {N_{\rm{BS}}}$.
In the uplink transmission, the received signal ${\bm{y}} \in \mathbb{C}^{N_{\rm{S}} \times 1}$ at the BS can be expressed as
\begin{equation}\label{eq:y}
\begin{array}{l}
{\bm{y}} = {{\bm{W}}^H}{\bm{HFs}} + {{\bm{W}}^H}{\bm{n}},
\end{array}
\end{equation}
where ${\bm{W}} = {{\bm{W}}_{\rm{RF}}}{{\bm{W}}_{\rm{BB}}} \in \mathbb{C}^{N_{\rm{BS}} \times N_{\rm{S}}}$ is the hybrid combiner.
${{\bm{W}}_{\rm{RF}} \in \mathbb{C}^{N_{\rm{BS}} \times N_{\rm{RF}}^{\rm{BS}}}}$ and ${{\bm{W}}_{\rm{BB}} \in \mathbb{C}^{N_{\rm{RF}}^{\rm{BS}} \times N_{\rm{S}}}}$ denote the analog and digital combiners, respectively.
${\bm{H}} \in \mathbb{C}^{N_{\rm{BS}} \times N_{\rm{MS}}}$ is the uplink channel matrix.
${\bm{F}} = {{\bm{F}}_{\rm{RF}}}{{\bm{F}}_{\rm{BB}}} \in \mathbb{C}^{N_{\rm{MS}} \times N_{\rm{S}}}$ is the hybrid precoder where ${{\bm{F}}_{\rm{RF}} \in \mathbb{C}^{N_{\rm{MS}} \times N_{\rm{RF}}^{\rm{MS}}}}$ and ${{\bm{F}}_{\rm{BB}} \in \mathbb{C}^{N_{\rm{RF}}^{\rm{MS}} \times N_{\rm{S}}}}$ denote the analog and digital precoders, respectively.
Note that every entry of ${{\bm{F}}_{\rm{RF}}}$ and ${{\bm{W}}_{\rm{RF}}}$ should satisfy the constraint of constant modulus, i.e. ${\left| {{\left[ {{\bm{F}}_{\rm{RF}}} \right]}_{m,n}} \right| = 1/\displaystyle\sqrt{N_{\rm{MS}}}}$ and ${\left| {{\left[ {{\bm{W}}_{\rm{RF}}} \right]}_{m,n}} \right| = 1/\displaystyle\sqrt{N_{\rm{BS}}}}$ for the $(m,n)$th elements of ${{\bm{F}}_{\rm{RF}}}$ and ${{\bm{W}}_{\rm{RF}}}$, respectively, since both ${{\bm{F}}_{\rm{RF}}}$ and ${{\bm{W}}_{\rm{RF}}}$ are realized by the analog RF phase shifters.
To guarantee the total transmit power constraint, the digital precoder ${\bm{F}}_{\rm{BB}}$ is further normalized as $\left\| {{{\bm{F}}_{\rm{RF}}}{{\bm{F}}_{\rm{BB}}}} \right\|_F^2 = N_{\rm{RF}}^{\rm{MS}}$.
${\bm{s}} \in \mathbb{C}^{N_{\rm{S}} \times 1}$ denotes the transmitted baseband signal from the MS, and ${\bm{n}} \in \mathbb{C}^{N_{\rm{BS}} \times 1}$ is the complex additive white Gaussian noise (AWGN) following the distribution ${\cal C}{\cal N}( {{\bm{0}},\sigma _n^2{\bm{I}}} )$ at the BS.

Due to the severe path loss for non-line-of-sight (NLOS) paths and thus the limited significant scatterers, the geometric mmWave channel with sparse multipath components is adopted \cite{{overview:Heath},{Hybrid:MIMO},{Auxiliary},{LimitedRF},{AdaptiveCS},{ACS},{OMP},{Subspace},{LSE},{Overlapped}}.
We consider only $L$ dominated paths corresponding to $L$ different scatterers contribute to the channel matrix $\bm{H}$, which can be written as
\begin{equation}\label{eq:channel}
\begin{array}{l}
{{\bm{H}} = \displaystyle\sqrt {\dfrac{{{N_{\rm{BS}}}{N_{\rm{MS}}}}}{L}} \displaystyle\sum\limits_{l = 1}^L {{\alpha _l}{{\bm{a}}_{\rm{BS}}}( {{\theta _l}} ){\bm{a}}_{\rm{MS}}^H( {{\varphi _l}} )}},
\end{array}
\end{equation}
where ${\alpha_l} \sim {\cal C}{\cal N}( {0,\sigma _\alpha ^2} )$ is the complex gain of the $l$th path, ${\theta _l}$ and ${\varphi _l}$ are the azimuth angles of AoA and AoD pair of the $l$th path, respectively.
Here, a typical uniform linear array (ULA) is considered at both BS and MS \cite{{overview:Heath},{Auxiliary},{LimitedRF},{AdaptiveCS},{ACS},{OMP},{Subspace},{LSE}}, so the steering vectors ${{\bm{a}}_{\rm{BS}}}( {{\theta _l}} )$ and ${{\bm{a}}_{\rm{MS}}}( {{\varphi _l}})$ associated with the $l$th path can be respectively written as
\begin{equation}\label{eq:steeringvec}
\begin{array}{l}
\hspace{-2mm}
{{{\bm{a}}_{\rm{BS}}}( {{\theta _l}} ) \!=\! \dfrac{1}{\displaystyle\sqrt {N_{\rm{BS}}}}{ \left[ {1,{e^{j2\pi \varDelta \sin ( {{\theta _l}} )}},\! \cdots \!,{e^{j2\pi ( {{N_{\rm{BS}}} - 1} )\varDelta \sin ( {{\theta _l}} )}}} \right]^T}},\\ [3.5mm]
\hspace{-2mm}
{{{\bm{a}}_{\rm{MS}}}( {{\varphi _l}} ) \!=\! \dfrac{1}{\displaystyle\sqrt {N_{\rm{MS}}}}{ \left[ {1,{e^{j2\pi \varDelta \sin ( {{\varphi _l}} )}},\! \cdots \!,{e^{j2\pi ( {{N_{\rm{MS}}} - 1} )\varDelta \sin ( {{\varphi _l}} )}}} \right]^T}},
\end{array}
\end{equation}
where $\varDelta  = d/\lambda$ denotes the normalized spacing of adjacent antennas, $\lambda$ is the wavelength, and $d$ is the spacing of adjacent antennas.
Furthermore, the mmWave channel matrix ${\bm{H}}$ can be rewritten in a more compact form as
\begin{align}\label{eq:H}
{\bm{H}} = {{\bm{A}}_{\rm{BS}}}{\bm{DA}}_{\rm{MS}}^H,
\end{align}
where ${{\bm{A}}_{\rm{BS}}} = \left[{{\bm{a}}_{\rm{BS}}}( {{\theta _1}} ), \cdots ,{{\bm{a}}_{\rm{BS}}}({\theta _L}) \right] \in \mathbb{C}^{N_{\rm{BS}} \times L}$, ${{\bm{A}}_{\rm{MS}}} = \left[{{\bm{a}}_{\rm{MS}}}( {{\varphi _1}} ), \cdots ,{{\bm{a}}_{\rm{MS}}}({\varphi _L}) \right] \in \mathbb{C}^{N_{\rm{MS}} \times L}$, and ${\bm{D}} = \displaystyle\sqrt {{N_{\rm{BS}}}{N_{\rm{MS}}}/L} \!\ {\rm{diag}}( {\bm{\alpha }} )$ is a diagonal matrix with ${\bm{\alpha }} = { \left[ {{\alpha _1}, \cdots ,{\alpha _L}} \right]^T}$.

\section{Proposed 2D Unitary ESPRIT Based Super-Resolution Channel Estimation Scheme}
In this section, we propose a 2D unitary ESPRIT based super-resolution channel estimation scheme for mmWave massive MIMO with hybrid precoding.
Specifically, the uplink training signals will be designed at first to estimate a low-dimensional effective channel having the same shift-invariance of array response as the high-dimensional mmWave MIMO channel matrix with low pilot overhead.
Then the super-resolution estimates of AoAs and AoDs are jointly obtained by exploiting the 2D unitary ESPRIT based channel estimation algorithm.
Moreover, the path gains are estimated by applying the LS estimator.
The high-dimensional mmWave MIMO channel therewith will be reconstructed according to the acquired AoAs, AoDs, and path gains.
Finally, the computational complexity of the proposed channel estimation scheme is compared with its conventional counterparts.

\vspace*{-2mm}
\subsection{Design of Training Signals for Uplink Channel Estimation}
To estimate the high-dimensional mmWave MIMO channel, we will first estimate the AoAs and AoDs with high accuracy.
Specifically, we will design the uplink training signals consisting of the analog RF part and digital baseband part at both BS and MS.
So that a low-dimensional effective channel having the same shift-invariance of array response as the high-dimensional mmWave channel can be acquired.
We begin by considering the uplink channel estimation in multiple time slots, where the received signal ${\bm{Y}} = \left[ {{{\bm{y}}_1}, \cdots ,{{\bm{y}}_{T_{\rm{MS}}}}} \right] \in \mathbb{C}^{N_{\rm{S}} \times {T_{\rm{MS}}}}$ in ${T_{\rm{MS}}}$ time slots or one time block can be expressed as
\begin{align}\label{eq:Y}
{\bm{Y}} = {{\bm{W}}^H}{\bm{HFS}} + {{\bm{W}}^H}{\bm{N}},
\end{align}
where ${\bm{S}} = \left[ {{{\bm{s}}_1}, \cdots ,{{\bm{s}}_{T_{\rm{MS}}}}} \right] \in \mathbb{C}^{N_{\rm{S}} \times {T_{\rm{MS}}}}$ denotes the transmitted pilot signal block, ${\bm{N}} = \left[ {{{\bm{n}}_1}, \cdots ,{{\bm{n}}_{T_{\rm{MS}}}}} \right] \in \mathbb{C}^{N_{\rm{BS}} \times {T_{\rm{MS}}}}$ is the AWGN in ${T_{\rm{MS}}}$ time slots, and we consider the channel matrix remains unchanged in the stage of channel estimation.
Furthermore, to improve the channel estimation performance, we further consider to exploit $N_{\rm{b}}^{\rm{T}}N_{\rm{b}}^{\rm{R}}$ time blocks jointly, so that the aggregated received signal ${\bm{\widetilde{Y}}} \in \mathbb{C}^{N_{\rm{b}}^{\rm{R}}{N_{\rm{S}}} \times N_{\rm{b}}^{\rm{T}}{T_{\rm{MS}}}}$ in $N_{\rm{b}}^{\rm{T}}$$N_{\rm{b}}^{\rm{R}}$ time blocks can be expressed as
\begin{align}\label{eq:Ytilde1}
{\bm{\widetilde{Y}}} = \left[\begin{array}{*{20}{c}}
{{{\bm{Y}}_{1,1}}}& \cdots &{{{\bm{Y}}_{1,N_{\rm{b}}^{\rm{T}}}}}\\
 \vdots & \ddots & \vdots \\
{{{\bm{Y}}_{N_{\rm{b}}^{\rm{R}},1}}}& \cdots &{{{\bm{Y}}_{N_{\rm{b}}^{\rm{R}},N_{\rm{b}}^{\rm{T}}}}}
\end{array} \right] = {{\bm{\widetilde{W}}}^H}{\bm{H\widetilde{F}\bar{S}}} + {{\bm{\bar{W}}}^H}{\bm{\widetilde{N}}},
\end{align}
where ${\bm{Y}}_{i,j} \in \mathbb{C}^{N_{\rm{S}} \times {T_{\rm{MS}}}}$, for $i = 1, \cdots ,N_{\rm{b}}^{\rm{R}}$ and $j = 1, \cdots ,N_{\rm{b}}^{\rm{T}}$, is the received signal in the $((i-1)N_{\rm{b}}^{\rm{T}}+j)$th time block,
\begin{align}\label{eq:FWtilde1}
\begin{array}{rcl}
{\bm{\widetilde{W}}} \! \! \! \! &=& \! \! \! \! \left[ {{{\bm{W}}_1}, \cdots , {{\bm{W}}_{N_{\rm{b}}^{\rm{R}}}}} \right] \in \mathbb{C}^{{N_{\rm{BS}}} \times N_{\rm{b}}^{\rm{R}}{N_{\rm{S}}}},\\ [2mm]
{\bm{\widetilde{F}}} \! \! \! \! &=& \! \! \! \! \left[ {{{\bm{F}}_1}, \cdots , {{\bm{F}}_{N_{\rm{b}}^{\rm{T}}}}} \right] \in \mathbb{C}^{{N_{\rm{MS}}} \times {N_{\rm{b}}^{\rm{T}}}{N_{\rm{S}}}},
\end{array}
\end{align}
are the aggregated hybrid combiner and precoder, respectively, and they will be designed later.
${\bm{\bar S}} = {\rm{diag}} \left[ {{\bm{S}}, \cdots ,{\bm{S}}} \right] \in \mathbb{C}^{{N_{\rm{b}}^{\rm{T}}}{N_{\rm{S}}} \times {N_{\rm{b}}^{\rm{T}}}{T_{\rm{MS}}}}$ is the aggregated pilot signal transmitted by the MS with ${N_{\rm{b}}^{\rm{T}}}$ identical pilot signal blocks ${\bm{S}}$ on the block diagonal, ${\bm{\bar W}} = {\rm{diag}} [ \bm{\widetilde{W}} ] \in \mathbb{C}^{{N_{\rm{b}}^{\rm{R}}}{N_{\rm{BS}}} \times {N_{\rm{b}}^{\rm{R}}}{N_{\rm{S}}}}$, and ${\bm{\widetilde{N}}} \in \mathbb{C}^{{N_{\rm{b}}^{\rm{R}}}{N_{\rm{BS}}} \times {N_{\rm{b}}^{\rm{T}}}{T_{\rm{MS}}}}$ is the aggregated AWGN matrix.
Thus, the total number of pilot overhead required for channel estimation is $T = {T_{\rm{MS}}}{N_{\rm{b}}^{\rm{R}}}{N_{\rm{b}}^{\rm{T}}}$.

As we have discussed in Section I, for mmWave MIMO with hybrid precoding, each baseband observation contains the signals from different antennas due to the RF phase shift network.
Hence, directly using the conventional ESPRIT algorithms can be difficult, since the shift-invariance of array response in these baseband observations is destroyed.
To solve this problem, we will design the aggregated precoder ${\bm{\widetilde{F}}}$ and combiner ${\bm{\widetilde{W}}}$, so that the shift-invariance of array response in the baseband observations can be preserved.
Particularly, we consider the aggregated precoder ${\bm{\widetilde{F}}}$ and combiner ${\bm{\widetilde{W}}}$ with the following forms, i.e.
\begin{equation}\label{eq:FWtilde2}
\begin{array}{rcl}
{\bm{\widetilde{F}}} \! \! \! \! &=& \! \! \! \! {\alpha _f} \left[ {\begin{array}{*{20}{c}}
{{\bm{I}}_{{N_{\rm{b}}^{\rm{T}}}{N_{\rm{S}}}}}\\
{{{\bm{O}}_{({N_{\rm{MS}}} - {N_{\rm{b}}^{\rm{T}}}{N_{\rm{S}}}) \times {N_{\rm{b}}^{\rm{T}}}{N_{\rm{S}}}}}}
\end{array}} \right],\\ [3mm]
{\bm{\widetilde{W}}} \! \! \! \! &=& \! \! \! \! {\alpha _w} \left[ {\begin{array}{*{20}{c}}
{{{\bm{I}}_{{N_{\rm{b}}^{\rm{R}}}{N_{\rm{S}}}}}}\\
{{{\bm{O}}_{({N_{\rm{BS}}} - {N_{\rm{b}}^{\rm{R}}}{N_{\rm{S}}}) \times {N_{\rm{b}}^{\rm{R}}}{N_{\rm{S}}}}}}
\end{array}} \right]{\rm{,}}
\end{array}
\end{equation}
where ${\alpha _f}$ and ${\alpha _w}$ are the scale factors for ${\bm{\widetilde{F}}}$ and ${\bm{\widetilde{W}}}$, respectively, to guarantee the constraints of constant modulus and power.
As a result, ${{\bm{\widetilde{W}}}^H}{\bm{H\widetilde{F}}}$ can be considered as the low-dimensional effective channel matrix ${\bm{\bar H}} \in \mathbb{C}^{{{N_{\rm{b}}^{\rm{R}}}{N_{\rm{S}}}} \times {{N_{\rm{b}}^{\rm{T}}}{N_{\rm{S}}}}}$ by substituting (\ref{eq:Hbar}), i.e.
\begin{equation} \label{eq:Hbar}
\begin{array}{rcl}
{\bm{\bar H}} \! \! \! \! &=& \! \! \! \! {{{\bm{\widetilde W}}}^H}{\bm{H\widetilde F}} \\ [3mm]
\! \! \! \! &=& \! \! \! \! {\alpha _w}{\alpha _f}\left[ {\begin{array}{*{20}{c}}
{{{\bm{H}}_{1,1}}}& \cdots &{{{\bm{H}}_{1,{{N_{\rm{b}}^{\rm{T}}}{N_{\rm{S}}}}}}}\\
 \vdots & \ddots & \vdots \\
{{{\bm{H}}_{{{N_{\rm{b}}^{\rm{R}}}{N_{\rm{S}}}},1}}}& \cdots &{{\bm{H}}_{{{N_{\rm{b}}^{\rm{R}}}{N_{\rm{S}}}},{{N_{\rm{b}}^{\rm{T}}}{N_{\rm{S}}}}}}
\end{array}} \right],
\end{array}
\end{equation}
where $\bm{H}_{m,n}$ represents the $(m, n)$th element of $\bm{H}$.
From (\ref{eq:Hbar}), we can observe that elements of the low-dimensional effective channel matrix ${\bm{\bar H}}$ come from the elements of the high-dimensional channel matrix $\bm{H}$ in first ${{N_{\rm{b}}^{\rm{R}}}{N_{\rm{S}}}}$ rows and first ${{N_{\rm{b}}^{\rm{T}}}{N_{\rm{S}}}}$ columns.
Hence, ${\bm{\bar H}}$ and $\bm{H}$ share the same shift-invariance of array response.
In this way, we can use ESPRIT algorithms to acquire the super-resolution estimates of AoAs and AoDs from the low-dimensional effective channel matrix ${\bm{\bar H}}$ instead of the original high-dimensional mmWave MIMO channel matrix $\bm{H}$ for the reduced pilot overhead.

Clearly, to facilitate the usage of ESPRIT algorithms, according to (\ref{eq:FWtilde1}), the analog and digital precoders $\left\{ {{\bm{F}}_{{\rm{RF}},j}} \right\}_{j = 1}^{N_{\rm{b}}^{\rm{T}}}$ and $\left\{ {{\bm{F}}_{{\rm{BB}},j}} \right\}_{j = 1}^{N_{\rm{b}}^{\rm{T}}}$ as well as the analog and digital combiners $\left\{ {{\bm{W}}_{{\rm{RF}},i}} \right\}_{i = 1}^{N_{\rm{b}}^{\rm{R}}}$ and $\left\{ {{\bm{W}}_{{\rm{BB}},i}} \right\}_{i = 1}^{N_{\rm{b}}^{\rm{R}}}$ should be well designed to guarantee (\ref{eq:FWtilde2}).
For clarity in what follows, we neglect the constraints of constant modulus of analog phase shifters network and total transmit power (namely, ${\alpha _f}$ and ${\alpha _w}$) for the precoder and combiner here.
To be specific, a unitary matrix ${{\bm{U}}_{N_{\rm{RF}}^{\rm{MS}}}} = \Big[ {{{\bm{u}}_1}, \cdots ,{{\bm{u}}_{N_{\rm{RF}}^{\rm{MS}}}}} \Big] \in \mathbb{C}^{{N_{\rm{RF}}^{\rm{MS}}} \times {N_{\rm{RF}}^{\rm{MS}}}}$ is considered as the set of the uplink training signals, such as a discrete Fourier transform (DFT) matrix, which has the orthogonality among different columns, i.e. ${\bm{u}}_m^H{{\bm{u}}_m} = N_{\rm{RF}}^{\rm{MS}}$ for ${m} = 1, \cdots ,N_{\rm{RF}}^{\rm{MS}}$, while ${\bm{u}}_m^H{{\bm{u}}_n} = 0$ for ${m \ne n}$.
Furthermore, we consider the digital precoder ${{\bm{F}}_{{\rm{BB}},j}}$, for $j = 1, \cdots ,N_{\rm{b}}^{\rm{T}}$, comes from the first $N_{\rm{S}}$ columns of ${{\bm{U}}_{N_{\rm{RF}}^{\rm{MS}}}}$, i.e.
\begin{align} \label{eq:FBB}
{{\bm{F}}_{{\rm{BB}},j}}= \left[ {{{\bm{u}}_1}, \cdots ,{{\bm{u}}_{{N_{\rm{S}}}}}} \right]
\end{align}
where ${N_{\rm{S}}} = N_{{\rm{RF}}}^{{\rm{MS}}} - 1$ is considered.
While for the analog precoder ${{\bm{F}}_{{\rm{RF}},j}} \in \mathbb{C}^{{N_{\rm{MS}}} \times N_{\rm{RF}}^{\rm{MS}}}$, we consider it has the following expression
\begin{align} \label{eq:FRF}
{{\bm{F}}_{{\rm{RF}},j}} = {\left[ {\bm{F}}_{{\rm{RF}},j}^1, {{\bm{F}}_{{\rm{BB}},j}}, {\bm{F}}_{{\rm{RF}},j}^2 \right]^H},
\end{align}
where ${\bm{F}}_{{\rm{RF}},j}^1 = \Big[ \underbrace {{{\bm{u}}_{N_{\rm{RF}}^{\rm{MS}}}}, \cdots ,{{\bm{u}}_{N_{\rm{RF}}^{\rm{MS}}}}}_{(j-1)N_{\rm{S}}} \Big]$ and ${\bm{F}}_{{\rm{RF}},j}^2 = \Big[ \underbrace {{{\bm{u}}_{N_{\rm{RF}}^{\rm{MS}}}}, \cdots ,{{\bm{u}}_{N_{\rm{RF}}^{\rm{MS}}}}}_{N_{\rm{MS}}-jN_{\rm{S}}} \Big]$ are composed of ${((j-1)N_{\rm{S}})}$ and ${(N_{\rm{MS}}-jN_{\rm{S}})}$ identical ${{\bm{u}}_{N_{\rm{RF}}^{\rm{MS}}}}$, respectively.
Based on the designed digital and analog precoders ${{\bm{F}}_{{\rm{BB}},j}}$ and ${{\bm{F}}_{{\rm{RF}},j}}$ in (\ref{eq:FBB}) and (\ref{eq:FRF}), we can have ${{\bm{F}}_j} = {{\bm{F}}_{{\rm{RF}},j}}{{\bm{F}}_{{\rm{BB}},j}}$.
Similarly, the digital combiner ${{\bm{W}}_{{\rm{BB}},i}}$, for $i = 1, \cdots ,N_{\rm{b}}^{\rm{R}}$, comes from the first $N_{\rm{S}}$ columns of ${{\bm{U}}_{N_{\rm{RF}}^{\rm{BS}}}}$, i.e. ${{\bm{W}}_{{\rm{BB}},i}} = \left[ {{{\bm{u}}_1}, \cdots ,{{\bm{u}}_{{N_{\rm{S}}}}}} \right]$.
The analog combiner ${{\bm{W}}_{{\rm{RF}},i}} = {\left[ {\bm{W}}_{{\rm{RF}},i}^1, {{\bm{W}}_{{\rm{BB}},i}}, {\bm{W}}_{{\rm{RF}},i}^2 \right]^H} \in \mathbb{C}^{{N_{\rm{BS}}} \times N_{\rm{RF}}^{\rm{BS}}}$, where ${\bm{W}}_{{\rm{RF}},i}^1 = \left[ {{\bm{u}}_{N_{\rm{RF}}^{\rm{BS}}}}, \cdots ,{{\bm{u}}_{N_{\rm{RF}}^{\rm{BS}}}} \right] \in \mathbb{C}^{{N_{\rm{RF}}^{\rm{BS}}} \times {((i-1)N_{\rm{S}})}}$ and ${\bm{W}}_{{\rm{RF}},i}^2 = \left[ {{\bm{u}}_{N_{\rm{RF}}^{\rm{MS}}}}, \cdots ,{{\bm{u}}_{N_{\rm{RF}}^{\rm{BS}}}} \right] \in \mathbb{C}^{{N_{\rm{RF}}^{\rm{BS}}} \times {(N_{\rm{BS}}-iN_{\rm{S}})}}$ are composed of ${((i-1)N_{\rm{S}})}$ and ${(N_{\rm{BS}}-iN_{\rm{S}})}$ identical ${{\bm{u}}_{N_{\rm{RF}}^{\rm{BS}}}}$, respectively.
Then, we can have ${{\bm{W}}_i} = {{\bm{W}}_{{\rm{RF}},i}}{{\bm{W}}_{{\rm{BB}},i}}$.
According to (\ref{eq:FWtilde1}), $N_{\rm{b}}^{\rm{T}}$ precoding matrices $\left\{ {\bm{F}_j} \right\}_{j = 1}^{N_{\rm{b}}^{\rm{T}}}$ and $N_{\rm{b}}^{\rm{R}}$ combining matrices $\left\{ {\bm{W}_i} \right\}_{i = 1}^{N_{\rm{b}}^{\rm{R}}}$ constitute the aggregated precoder ${\bm{\widetilde{F}}}$ and the aggregated combiner ${\bm{\widetilde{W}}}$, respectively.
Finally, we have
\begin{equation}\label{eq:FWtilde3}
\begin{array}{rcl}
{\bm{\widetilde{F}}} \! \! \! \! &=& \! \! \! \!\left[ {{{\bm{F}}_1}, \cdots , {{\bm{F}}_{N_{\rm{b}}^{\rm{T}}}}} \right] = {\alpha _f}\left[ {\begin{array}{*{20}{c}}
{{\bm{I}}_{{N_{\rm{b}}^{\rm{T}}}{N_{\rm{S}}}}}\\
{{{\bm{O}}_{({N_{\rm{MS}}} - {N_{\rm{b}}^{\rm{T}}}{N_{\rm{S}}}) \times {N_{\rm{b}}^{\rm{T}}}{N_{\rm{S}}}}}}
\end{array}} \right],\\ [3.5mm]
{\bm{\widetilde{W}}} \! \! \! \! &=& \! \! \! \!\left[ {{{\bm{W}}_1}, \cdots , {{\bm{W}}_{N_{\rm{b}}^{\rm{R}}}}} \right] = {\alpha _w}\left[ {\begin{array}{*{20}{c}}
{{{\bm{I}}_{{N_{\rm{b}}^{\rm{R}}}{N_{\rm{S}}}}}}\\
{{{\bm{O}}_{({N_{\rm{BS}}} - {N_{\rm{b}}^{\rm{R}}}{N_{\rm{S}}}) \times {N_{\rm{b}}^{\rm{R}}}{N_{\rm{S}}}}}}
\end{array}} \right]{\rm{.}}
\end{array}
\end{equation}
As a consequence, (\ref{eq:Ytilde1}) can be further written as
\begin{equation}\label{eq:Ytilde2}
\begin{array}{rcl}
{\bm{\widetilde{Y}}} \! \! \! \! &=& \! \! \! \! {\alpha _w}{\alpha _f}\left[ {\begin{array}{*{20}{c}}
{{\bm{I}}_{{N_{\rm{b}}^{\rm{R}}}{N_{\rm{S}}}}}&{{{\bm{O}}_{({N_{\rm{BS}}} - {N_{\rm{b}}^{\rm{R}}}{N_{\rm{S}}}) \times {N_{\rm{b}}^{\rm{R}}}{N_{\rm{S}}}}}}
\end{array}} \right]{\bm{H}} \\ [1mm]
\! \! \! \! &\ & \! \! \! \! \left[ {\begin{array}{*{20}{c}}
{{\bm{I}}_{{N_{\rm{b}}^{\rm{T}}}{N_{\rm{S}}}}}\\
{{{\bm{O}}_{({N_{\rm{MS}}} - {N_{\rm{b}}^{\rm{T}}}{N_{\rm{S}}}) \times {N_{\rm{b}}^{\rm{T}}}{N_{\rm{S}}}}}}
\end{array}} \right]{\bm{\bar{S}}}+{{\bm{\bar{W}}}^H}{\bm{\widetilde{N}}} \\ [4.5mm]
\! \! \! \! &=& \! \! \! \! {\bm{\bar H}}{\bm{\bar{S}}}+{{\bm{\bar{W}}}^H}{\bm{\widetilde{N}}}.
\end{array}
\end{equation}

To accurately acquire the low-dimensional effective channel matrix ${\bm{\bar H}}$ having the same shift-invariance of array response as the high-dimensional channel matrix $\bm{H}$ from the aggregated received signal ${\bm{\widetilde{Y}}}$, we can use the LS estimator to obtain its estimate as ${\widehat{\bm{\bar H}}} = {\bm{\widetilde{Y}}}{\bm{\bar{S}}}^\dagger = {\bm{\widetilde{Y}}}{{\bm{\bar S}}^H}{\left( {{\bm{\bar S}}{{{\bm{\bar S}}}^H}} \right)^{-1}}$.
For convenience, here we consider the transmit pilot signal block ${\bm{S}}$ as a unitary matrix, which has the perfect autocorrelation property (i.e. ${\bm{S}}{{\bm{S}}^H} = {N_{\rm{S}}}{{\bm{I}}_{{N_{\rm{S}}}}}$ with $T_{\rm{MS}} = N_{\rm{S}}$).
In this way, the low-dimensional effective channel matrix can be written as ${\bm{\bar H}} = {\bm{\widetilde Y}}{{\bm{\bar S}}^H}/{N_{\rm{S}}}$.

Based on above design, we can acquire ${\bm{\bar H}}$ preserving the shift-invariance of array response.
This motivates us to exploit the ESPRIT algorithm to estimate the AoAs/AoDs jointly, which will be illustrated in the following subsection.

\subsection{2D Unitary ESPRIT Based Channel Estimation Algorithm}
To jointly obtain the super-resolution estimates of AoAs and AoDs, we propose a 2D unitary ESPRIT based channel estimation algorithm at the receiver, which includes the following several main steps and is summarized in \textbf{Algorithm 1}.
Note that for the low-dimensional effective channel matrix ${\bm{\bar H}}$ in (\ref{eq:Hbar}), we consider it has the size of ${N_{\rm{R}}} \times {N_{\rm{T}}}$ for convenience.

\subsubsection{Construct Hankel Matrix and Extend Data}
To alleviate the influence of coherent signals caused by multiple AoAs or AoDs close to each other, we consider the \emph{spatial smoothing} and the \emph{forward backward averaging} techniques in \cite{{JADE:SIT}}.
By leveraging these two techniques, we can take full advantage of obtained data, and acquire a robust AoAs and AoDs estimation to mitigate the performance loss due to rank-deficiency of the data matrix when multiple AoAs or AoDs are close to each other.
Specifically, we introduce integers $m_1$ and $m_2$ as the stacking parameters, where $2 \le {m_1} \le {N_{\rm{T}}}$ and $1 \le {m_2} \le {N_{\rm{R}}} - 1$.
Meanwhile, for $1 \le {i} \le {m_2}$ and $1 \le {j} \le {m_1}$, we define the left/right-shifted matrix as ${{\bm{\bar H}}^{\left( {i,j} \right)}} \in \mathbb{C}^{({N_{\rm{R}}} - {m_2} + 1) \times ({N_{\rm{T}}} - {m_1} + 1)}$, which is a submatrix of ${\bm{\bar H}}$, and it can be written as
\begin{align}\label{eq:Hbarij}
\hspace{-2mm}
{{\bm{\bar H}}^{\left( {i,j} \right)}} = \left[ {\begin{array}{*{20}{c}}
{{{{\bm{\bar H}}}_{i,j}}}& \cdots &{{{{\bm{\bar H}}}_{i,{N_{\rm{T}}} - {m_1} + j}}}\\
 \vdots & \ddots & \vdots \\
{{{{\bm{\bar H}}}_{{N_{\rm{R}}} - {m_2} + i,j}}}& \cdots &{{{{\bm{\bar H}}}_{{N_{\rm{R}}} - {m_2} + i,{N_{\rm{T}}} - {m_1} + j}}}
\end{array}} \right],
\end{align}
where ${{\bm{\bar H}}_{i,j}}$ represents the $(i,j)$th element of ${\bm{\bar H}}$ in (\ref{eq:Hbar}).
Furthermore, we can construct a $Hankel$ matrix $\bm{{\cal H}} \in \mathbb{C}^{{m_1}({N_{\rm{R}}} - {m_2} + 1){\rm{ }} \times {m_2}({N_{\rm{T}}} - {m_1} + 1)}$ as
\begin{align}\label{eq:Hankel}
{\bm{{\cal H}}} = \left[ {\begin{array}{*{20}{c}}
{{{\bm{\bar {H}}}}^{\left( {1,1} \right)}}& \cdots &{{{{\bm{\bar {H}}}}^{\left( {{m_2},1} \right)}}}\\
 \vdots & \ddots & \vdots \\
{{{{\bm{\bar {H}}}}^{\left( {1,{m_1}} \right)}}}& \cdots &{{{{\bm{\bar {H}}}}^{\left( {{m_2},{m_1}} \right)}}}
\end{array}} \right].
\end{align}

According to \cite{{JADE:SIT}}, the extend data matrix ${{\bm{{\cal H}}}_{\rm{e}}} \in \mathbb{C}^{{m_1}({N_{\rm{R}}} - {m_2} + 1){\rm{ }} \times 2{m_2}({N_{\rm{T}}} - {m_1} + 1)}$ can be written as
\begin{align}\label{eq:Hankele}
{{\bm{{\cal H}}}_{\rm{e}}} = \left[ {\begin{array}{*{20}{c}}
{\bm{{\cal H}}}&{{{\bm{J}}_{{m_1}({N_{\rm{R}}} - {m_2} + 1)}}}
\end{array}{{\bm{{\cal H}}}^*}} \right].
\end{align}

\subsubsection{Real Processing}
To reduce the computational complexity in the following steps, ${{\bm{{\cal H}}}_{\rm{e}}}$ is further transformed into the real matrix by left-multiplying and right-multiplying a transform matrix ${{\bm{T}}_{\rm{L}}}$ and ${{\bm{T}}_{\rm{R}}}$ to ${{\bm{{\cal H}}}_{\rm{e}}}$, respectively, so that the corresponding eigenvalues are real \cite{{2Du:esprit},{JADE}}.
Particularly, this manipulation can be expressed as ${\bm{{\cal H}}_{{\rm{e}},{\rm{R}}}} = {{\bm{T}}_{\rm{L}}}{\bm{{\cal H}}_{\rm{e}}}{{\bm{T}}_{\rm{R}}} \in \mathbb{R}^{{m_1}({N_{\rm{R}}} - {m_2} + 1) \times 2{m_2}({N_{\rm{T}}} - {m_1} + 1)}$, and the transformation matrices ${{\bm{T}}_{\rm{L}}}$ and ${{\bm{T}}_{\rm{R}}}$ can be respectively expressed as
\begin{align}\label{eq:TLTR}
\begin{array}{rcl}
{{\bm{T}}_{\rm{L}}} \! \! \! \! &=& \! \! \! \! {\bm{Q}}_{m1}^H \otimes {\bm{Q}}_{{N_{\rm{R}}} - {m_2} + 1}^H{\rm{,}}\\ [2mm]
{{\bm{T}}_{\rm{R}}} \! \! \! \! &=& \! \! \! \! \left[ {\begin{array}{*{20}{c}}
{{{\bm{I}}_{{m_2}({N_{\rm{T}}} - {m_1} + 1)}}}&{j{{\bm{I}}_{{m_2}({N_{\rm{T}}} - {m_1} + 1)}}}\\
{{{\bm{I}}_{{m_2}({N_{\rm{T}}} - {m_1} + 1)}}}&{ - j{{\bm{I}}_{{m_2}({N_{\rm{T}}} - {m_1} + 1)}}}
\end{array}} \right],
\end{array}
\end{align}
where ${\bm{Q}}$ is a particular $left$ ${\bm{J}}$-$real$ matrix according to \cite{{2Du:esprit}}, satisfying ${\bm{J}}{{\bm{Q}}^*} = {\bm{Q}}$, and it has the sparse and unitary properties, defined as
\begin{align}\label{eq:Q}
\begin{array}{rcl}
{{\bm{Q}}_{2n}} \! \! \! \! &=& \! \! \! \! \dfrac{1}{{\displaystyle\sqrt 2 }}\left[ {\begin{array}{*{20}{c}}
{{{\bm{I}}_n}}&{j{{\bm{I}}_n}}\\
{{{\bm{J}}_n}}&{ - j{{\bm{J}}_n}}
\end{array}} \right],\\ [4mm]
{{\bm{Q}}_{2n + 1}} \! \! \! \! &=& \! \! \! \! \dfrac{1}{{\displaystyle\sqrt 2 }}\left[ {\begin{array}{*{20}{c}}
{{{\bm{I}}_n}}&{{{\bm{0}}_n}}&{j{{\bm{I}}_n}}\\
{{\bm{0}}_n^T}&{\displaystyle\sqrt 2 }&{{\bm{0}}_n^T}\\
{{{\bm{J}}_n}}&{{{\bm{0}}_n}}&{ - j{{\bm{J}}_n}}
\end{array}} \right].
\end{array}
\end{align}

\subsubsection{Rank Reduction}
In the absence of noise, ${\bm{{\cal H}}_{{\rm{e}},{\rm{R}}}}$ has only $L \ll {\rm{min}}\{{m_1}({N_{\rm{R}}}-{m_2}+1), 2{m_2}({N_{\rm{T}}}-{m_1}+1)\}$ effective rank.
However, in the presence of noise, such low rank property of ${\bm{{\cal H}}_{{\rm{e}},{\rm{R}}}}$ is destroyed.
To mitigate the noise, the singular value decomposition (SVD) of ${\bm{{\cal H}}_{{\rm{e}},{\rm{R}}}}$, i.e. ${\bm{{\cal H}}_{{\rm{e}},{\rm{R}}}} = {\bm{U\varSigma }}{{\bm{V}}^H}$, will be used to distinguish the signal subspace and noise subspace.
In order to extract the information of AoAs and AoDs in the real matrix ${\bm{{\cal H}}_{{\rm{e}},{\rm{R}}}}$, we subsequently take the first $L$ columns of the left singular matrix $\bm{U}$, denoted as ${{\bm{\widehat{U}}}_{{\rm{e}},{\rm{R}}}} \in \mathbb{R}^{{{m_1}({N_{\rm{R}}} - {m_2} + 1)} \times L}$, to approximate the dominant $L$-dimensional column span of ${\bm{{\cal H}}_{{\rm{e}},{\rm{R}}}}$.

\begin{algorithm}[t]
\caption{ 2D Unitary ESPRIT Based Channel Estimation Algorithm} \label{Algorithm:1}
\begin{algorithmic}[1]
\REQUIRE ~~\\
The low-dimensional effective channel matrix ${\bm{\bar H}}$, the stacking parameters $m_1$ and $m_2$, and the number of paths $L$;
\ENSURE ~~\\
The estimated AoAs $\big\{\hat \theta _l  \big\}_{l = 1}^L$ and AoDs $\big\{\hat \varphi _l \big\}_{l = 1}^L$ of channel.
\STATE Construct the $Hankel$ matrix ${\bm{{\cal H}}}$ as shown in (\ref{eq:Hankel});
\label{step:1}
\STATE Obtain the extended matrix ${{\bm{{\cal H}}}_{\rm{e}}}$ as shown in (\ref{eq:Hankele});
\label{step:2}
\STATE Implement the real processing to achieve ${\bm{{\cal H}}_{{\rm{e}},{\rm{R}}}}$ in the real domain expressed as ${\bm{{\cal H}}_{{\rm{e}},{\rm{R}}}} = {{\bm{T}}_{\rm{L}}}{\bm{{\cal H}}_{\rm{e}}}{{\bm{T}}_{\rm{R}}}$, where ${{\bm{T}}_{\rm{L}}}$ and ${{\bm{T}}_{\rm{R}}}$ are shown in (\ref{eq:TLTR});
\label{step:3}
\STATE Let ${\bm{{\cal H}}_{{\rm{e}},{\rm{R}}}} = {\bm{U\varSigma }}{{\bm{V}}^H}$ and take the first $L$ columns of the left singular matrix $\bm{U}$, denoted as ${{\bm{\widehat U}}_{{\rm{e}},{\rm{R}}}}$;
\label{step:4}
\STATE Diagonalize to jointly estimate the AoAs/AoDs pairs ${\hat \theta _l}$ and ${\hat \varphi _l}$ according to (\ref{eq:thetaphi}) from the EVD of matrix $\bm{\varPsi}$ in (\ref{eq:Psi}), where ${\bm{\varPsi }} = {\bm{T\varLambda }}{{\bm{T}}^{ - 1}}$ with ${\bm{\widetilde \varPhi }} = {\rm{Re}}\left\{ {\bm{\varLambda }} \right\}$, ${\bm{\widetilde \varTheta }} = {\rm{Im}}\left\{ {\bm{\varLambda }} \right\}$.
\label{step:5}
\end{algorithmic}
\end{algorithm}

\subsubsection{Joint Diagonalization}
According to \cite{{2Du:esprit}} and \cite{{JADE}}, for a certain non-singular matrix ${\bm{T}} \in \mathbb{R}^{L \times L}$, we can obtain
\begin{align}\label{eq:JD1}
\begin{array}{rcl}
{{\bm{E}}_{\theta ,{\rm{R}}}}{{{\bm{\widehat U}}}_{{\rm{e}},{\rm{R}}}}{\bm{T\widetilde {\bm{\varTheta}} }} \! \! \! \! &=& \! \! \! \! {{\bm{E}}_{\theta ,{\rm{I}}}}{{{\bm{\widehat U}}}_{{\rm{e}},{\rm{R}}}}{\bm{T}},\\ [2mm]
{{\bm{E}}_{\varphi ,{\rm{R}}}}{{{\bm{\widehat U}}}_{{\rm{e}},{\rm{R}}}}{\bm{T\widetilde {\bm{\varPhi}} }} \! \! \! \! &=& \! \! \! \! {{\bm{E}}_{\varphi ,{\rm{I}}}}{{{\bm{\widehat U}}}_{{\rm{e}},{\rm{R}}}}{\bm{T}},
\end{array}
\end{align}
where ${{\bm{E}}_{\theta ,{\rm{R}}}} = {\rm{Re}}\left\{ {{{\bm{E}}_\theta }} \right\}$, ${{\bm{E}}_{\theta ,{\rm{I}}}} = {\rm{Im}}\left\{ {{{\bm{E}}_\theta }} \right\}$, ${{\bm{E}}_{\varphi ,{\rm{R}}}} = {\rm{Re}}\left\{ {{{\bm{E}}_\varphi }} \right\}$, and ${{\bm{E}}_{\varphi ,{\rm{I}}}} = {\rm{Im}}\left\{ {{{\bm{E}}_\varphi }} \right\}$ with
\begin{align}
\begin{array}{rcl}
{{\bm{E}}_\theta } \! \! \! \! &=& \! \! \! \! {{\bm{I}}_{{m_1}}} \otimes \left( {{\bm{Q}}_{{N_{\rm{R}}} - {m_2}}^H\left[ {\begin{array}{*{20}{c}}
{\bm{0}}&{{{\bm{I}}_{{N_{\rm{R}}} - {m_2}}}}
\end{array}} \right]{{\bm{Q}}_{{N_{\rm{R}}} - {m_2} + 1}}} \right),\nonumber \\ [2mm]
{{\bm{E}}_\varphi } \! \! \! \! &=& \! \! \! \! \left( {{\bm{Q}}_{{m_1} - 1}^H\left[ {\begin{array}{*{20}{c}}
{\bm{0}}&{{{\bm{I}}_{{m_1} - 1}}}
\end{array}} \right]{{\bm{Q}}_{{m_1}}}} \right) \otimes {{\bm{I}}_{{N_{\rm{R}}} - {m_2} + 1}},
\end{array}
\end{align}
respectively.
In (\ref{eq:JD1}), ${\bm{\widetilde \varTheta }} \in \mathbb{R}^{L \times L}$ and ${\bm{\widetilde \varPhi }} \in \mathbb{R}^{L \times L}$ are diagonal matrices, and they can be expressed as
\begin{align}\label{eq:ThetaPhi}
\begin{array}{rcl}
{\bm{\widetilde \varTheta }} \! \! \! \! &=& \! \! \! \! {\rm{diag}} ( {{{\tilde \theta }_1}, \cdots ,{{\tilde \theta }_L}} ),\\ [2mm]
{\bm{\widetilde \varPhi }} \! \! \! \! &=& \! \! \! \! {\rm{diag}} \left( {{{\tilde \varphi }_1}, \cdots ,{{\tilde \varphi }_L}} \right),\\ [2mm]
\end{array}
\end{align}
where
\begin{align}\label{eq:thetaphi}
\begin{array}{rcl}
{{\tilde \theta }_l} \! \! \! &=& \! \! \! \tan ( {\pi \varDelta \sin ( {{{\hat \theta }_l}} )} ),\\ [3mm]
{{\tilde \varphi }_l} \! \! \! &=& \! \! \! \tan \left( {\pi \varDelta \sin \left( {{{\hat \varphi }_l}} \right)} \right),
\end{array}
\end{align}
for ${l} = 1, \cdots ,L$, respectively.
Since $\bm{T}$ is an invertible square matrix, (\ref{eq:JD1}) can be further written as
\begin{align}\label{eq:JD2}
\begin{array}{rcl}
{\bm{T\widetilde \varTheta }}{{\bm{T}}^{ - 1}} \! \! \! \! &=& \! \! \! \! {\left( {{{\bm{E}}_{\theta ,{\rm{R}}}}{{{\bm{\widehat U}}}_{{\rm{e}},{\rm{R}}}}} \right)^\dagger }{{\bm{E}}_{\theta ,{\rm{I}}}}{{{\bm{\widehat U}}}_{{\rm{e}},{\rm{R}}}},\\ [2mm]
{\bm{T\widetilde \varPhi }}{{\bm{T}}^{ - 1}} \! \! \! \! &=& \! \! \! \! {\left( {{{\bm{E}}_{\varphi ,{\rm{R}}}}{{{\bm{\widehat U}}}_{{\rm{e}},{\rm{R}}}}} \right)^\dagger }{{\bm{E}}_{\varphi ,{\rm{I}}}}{{{\bm{\widehat U}}}_{{\rm{e}},{\rm{R}}}}.
\end{array}
\end{align}
According to (\ref{eq:thetaphi}), ${\bm{T\widetilde \varTheta }}{{\bm{T}}^{ - 1}}$ and ${\bm{T}{\widetilde {\bm{\varPhi}} }}{{\bm{T}}^{ - 1}}$ can be jointly diagonalized, which can be expressed as
\begin{align}\label{eq:Psi}
\begin{array}{rcl}
{\bm{\varPsi }}\! \! \! \! &=& \! \! \! \!\left( {{\bm{T\widetilde \varPhi }}{{\bm{T}}^{ - 1}}} \right) + j\left( {{\bm{T\widetilde \varTheta }}{{\bm{T}}^{ - 1}}} \right)\\ [2mm]
\! \! \! \! &=& \! \! \! \! {\left( {{{\bm{E}}_{\varphi ,{\rm{R}}}}{{{\bm{\widehat U}}}_{{\rm{e}},{\rm{R}}}}} \right)^\dagger }{{\bm{E}}_{\varphi ,{\rm{I}}}}{{{\bm{\widehat U}}}_{{\rm{e}},{\rm{R}}}} + j{\left( {{{\bm{E}}_{\theta ,{\rm{R}}}}{{{\bm{\widehat U}}}_{{\rm{e}},{\rm{R}}}}} \right)^\dagger }{{\bm{E}}_{\theta ,{\rm{I}}}}{{{\bm{\widehat U}}}_{{\rm{e}},{\rm{R}}}}.
\end{array}
\end{align}
Since ${\left( {{{\bm{E}}_{\theta ,{\rm{R}}}}{{{\bm{\widehat U}}}_{{\rm{e}},{\rm{R}}}}} \right)^\dagger }{{\bm{E}}_{\theta ,{\rm{I}}}}{{{\bm{\widehat U}}}_{{\rm{e}},{\rm{R}}}}$ and ${\left( {{{\bm{E}}_{\varphi ,{\rm{R}}}}{{{\bm{\widehat U}}}_{{\rm{e}},{\rm{R}}}}} \right)^\dagger }{{\bm{E}}_{\varphi ,{\rm{I}}}}{{{\bm{\widehat U}}}_{{\rm{e}},{\rm{R}}}}$ have the same eigenvectors, i.e. the column vectors of $\bm{T}$, ${\tilde \theta _l}$ and ${\tilde \varphi _l}$ corresponding to the same eigenvector in  ${\bm{\varPsi }}$, for ${l} = 1, \cdots ,L$, are associated with the same path.
That is to say, the estimated AoAs/AoDs pairs ${\hat \theta _l}$ and ${\hat \varphi _l}$ can be naturally paired by exploiting a complex eigenvalue decomposition (EVD), given by ${\bm{\varPsi }} = {\bm{T\varLambda }}{{\bm{T}}^{ - 1}}$, where ${\bm{\varLambda }} = {\bm{\widetilde \varPhi }} + j{\bm{\widetilde \varTheta }}$, ${\bm{\widetilde \varPhi }} = {\rm{Re}}\left\{ {\bm{\varLambda }} \right\}$, and ${\bm{\widetilde \varTheta }} = {\rm{Im}}\left\{ {\bm{\varLambda }} \right\}$ in (\ref{eq:Psi}).
Finally, according to (\ref{eq:ThetaPhi}) and (\ref{eq:thetaphi}), we can obtain the paired super-resolution estimates of AoAs and AoDs, i.e. $\big\{\hat \theta _l  \big\}_{l = 1}^L$ and $\big\{\hat \varphi _l \big\}_{l = 1}^L$.

\subsection{Reconstruct High-Dimensional mmWave MIMO Channel}
In this subsection, the high-dimensional mmWave MIMO channel will be reconstructed according to the obtained AoAs $\big\{\hat \theta _l  \big\}_{l = 1}^L$ and AoDs $\big\{\hat \varphi _l \big\}_{l = 1}^L$ in (\ref{eq:thetaphi}).
First of all, we reconstruct the matrices ${{\bm{\widehat A}}_{\rm{BS}}}$ and ${{\bm{\widehat A}}_{\rm{MS}}}$ according to the steering vectors ${{\bm{a}}_{\rm{BS}}}( {{\hat \theta _l}} )$ and ${{\bm{a}}_{\rm{MS}}}( {{\hat \varphi _l}} )$ in (\ref{eq:steeringvec}).
Then, based on (\ref{eq:H}) and (\ref{eq:Hbar}), we have the expression ${\bm{\bar H}} = {{\bm{\widetilde W}}^H}{{\bm{\widehat A}}_{\rm{BS}}}{\bm{D\widehat A}}_{\rm{MS}}^H{\bm{\widetilde F}}$ with $\bm{D} = {{\rm{diag}}( {\bm{d}} )}$, where ${\bm{d}} = \sqrt {{N_{\rm{BS}}}{N_{\rm{MS}}}/L} \!\  { \left[ {{\alpha _1}, \cdots ,{\alpha _L}} \right]^T}$.
To acquire the associated path gains, we vectorize the low-dimensional effective channel matrix ${\bm{\bar H}}$ as
\begin{align}\label{eq:hbar}
{\bm{\bar h}} = \left[ {{{\left( {{\bm{\widehat A}}_{\rm{MS}}^H{\bm{\widetilde F}}} \right)}^T} \odot \left( {{{{\bm{\widetilde W}}}^H}{{{\bm{\widehat A}}}_{\rm{BS}}}} \right)} \right]{\bm{d}} = {\bm{{Z} d}},
\end{align}
where ${\bm{\bar h}} = {\rm{vec}}\left( {{\bm{\bar H}}} \right)$, ${\bm{{Z}}} = {\left( {{\bm{\widehat A}}_{\rm{MS}}^H{\bm{\widetilde F}}} \right)^T} \odot \left( {{{{\bm{\widetilde W}}}^H}{{{\bm{\widehat A}}}_{\rm{BS}}}} \right)$, and we use the identity ${{\rm{vec}}\left( {{\bm{ABC}}} \right)} = \left( {{{\bm{C}}^T} \odot {\bm{A}}} \right){\bm{b}}$ with ${\bm{B}} = {\rm{diag}}\left( {\bm{b}} \right)$.
Using the LS estimator, we can obtain the LS solution ${\bm{\widehat d}}$, i.e.
\begin{align}\label{eq:dhat}
{\bm{\widehat d}} = \arg \mathop {\min }\limits_{\bm{d}} {\left\| {{\bm{\bar h}} - {\bm{{Z} d}}} \right\|_2} = {{\bm{{Z}}}^\dagger }{\bm{\bar h}} = {\left( {{{\bm{{Z}}}^H}{\bm{{Z}}}} \right)^{ - 1}}{{\bm{{Z}}}^H}{\bm{\bar h}}.
\end{align}

Finally, according to the obtained steering vector matrices ${{\bm{\widehat A}}_{\rm{BS}}}$, ${{\bm{\widehat A}}_{\rm{MS}}}$, and the gain of paths ${\bm{\widehat d}}$ above, we can reconstruct the high-dimensional mmWave MIMO channel as ${{\bm{\widehat H}} = {{\bm{\widehat A}}_{\rm{BS}}}{\rm{diag}}\Big( {{\bm{\widehat d}}} \Big){\bm{\widehat A}}_{\rm{MS}}^H}$.

\subsection{Analysis of Computational Complexity}
From \textbf{Algorithm 1}, it can be observed that the main computational complexity comes from the SVD in Step \ref{step:4} as well as the matrix inversion and EVD in Step \ref{step:5}.
While the computational complexity of the rest implementations, such as the basic matrix multiplications, can be negligible.
Specifically, for Step \ref{step:4}, the computational complexity of partial SVD taking the first $L$ columns of left singular matrix $\bm{U}$ is the order of ${\cal O}\left( {{m_1}\left( {{N_{\rm{R}}} - {m_2} + 1} \right){L^2}} \right)$ \cite{{Matrix}}.
While for Step \ref{step:5}, the computational complexity of the matrix inversion operations of both the real and imaginary parts in (\ref{eq:JD2}) and the following EVD are ${\cal O}\left( {{L^3}} \right)$ \cite{{Matrix}}, which can be small since the number of dominated paths $L$ is small due to the limited number of scatterers over mmWave channels.
Hence, the main computational cost of \textbf{Algorithm 1} lies in Step \ref{step:4}, where the computational complexity is ${\cal O}\left( {{m_1}\left( {{N_{\rm{R}}} - {m_2} + 1} \right){L^2}} \right)$.

In this subsection, we also consider the ACS-based channel estimation scheme \cite{{ACS}} and OMP-based channel estimation scheme \cite{{OMP}} for comparison.
For the ACS-based channel estimation scheme, the computational complexity is ${\cal O}\left( {2LN_{\rm{BS}}^3{{\log }_K}\left( {G_{\rm{ACS}}/L} \right)} \right)$ \cite{{Matrix}}, where $K$ is the number of beamforming vectors in each stage, and $G_{\rm{ACS}}$ is the number of uniform grid points.
For the OMP-based channel estimation scheme, the main computational costs lie in the correlation operation and matrix inversion operation.
Hence, its computational complexity is ${\cal O}\left( {N_{\rm{T}}^{\rm{Beam}}N_{\rm{R}}^{\rm{Beam}}G_{\rm{OMP}}^2 + {{\left| {{\cal I}_t} \right|}^4}} \right)$ \cite{{OMP},{Matrix}}, where $N_{\rm{T}}^{\rm{Beam}}$ and $N_{\rm{R}}^{\rm{Beam}}$ are the numbers of transmit and receive pilot beam patterns at the MS and BS, respectively.
$G_{\rm{OMP}}$ is the size of quantized grids of virtual AoAs/AoDs, and ${{\cal I}_t}$ is the cardinality of index set (here ${\left| {{{\cal I}_t}} \right|}$ is equal to the number of iterations for channel estimation).
It should be pointed out that the matrix inversion operation with the computational complexity of ${\cal O}\left( {{{\left| {{{\cal I}_t}} \right|}^4}} \right)$ will dominate the computational complexity of the OMP-based channel estimation scheme when the number of iterations ${\left| {{{\cal I}_t}} \right|}$ becomes very large, hundreds for instance.

Based on the analysis above, it can be observed that the computational complexity of the proposed scheme is proportional to $N_{\rm{R}}$, namely the number of rows of the low-dimensional effective channel matrix ${\bm{\bar H}}$.
By contrast, those of the ACS-based channel estimation scheme and the OMP-based channel estimation scheme are proportional to $N_{\rm{BS}}^3$ and ${{{\left| {{{\cal I}_t}} \right|}^4}}$ (or $G_{\rm{OMP}}^2$), respectively.
Hence the computational complexity of the 2D unitary ESPRIT based channel estimation scheme is lower compared with the ACS-based channel estimation scheme and the OMP-based channel estimation scheme.
In Section IV, the computational complexity among three different schemes will be further compared in the specific simulations.

\section{Simulation Results}
In this section, we will investigate the performance of the proposed 2D unitary ESPRIT based channel estimation scheme by comparing it with the ACS-based channel estimation scheme \cite{{ACS}} and the OMP-based channel estimation scheme \cite{{OMP}}.
We consider the simulation parameters shown as follows.
Specifically, $N_{\rm{BS}} = N_{\rm{MS}} = 64$, $N_{\rm{RF}} = N_{\rm{RF}}^{\rm{BS}} = N_{\rm{RF}}^{\rm{MS}} =4$, $T_{\rm{MS}} = N_{\rm{S}} = 3$, ${N_{\rm{b}}^{\rm{R}}}={N_{\rm{b}}^{\rm{T}}}=10$, $m_1 = m_2 = 13$, $\varDelta = 1/2$ (namely, $d = \lambda/2$), $\sigma _\alpha ^2 = 1$, and $\left\{ \theta _l  \right\}_{l = 1}^L$ and $\left\{ \varphi _l \right\}_{l = 1}^L$ follow the uniform distribution $\left[ { - \dfrac{\pi }{3},\dfrac{\pi }{3}} \right]$. 
The metrics for performance evaluation include the normalized mean square error (NMSE), defined as
\begin{align}\label{eq:NMSE}
{\rm{NMSE}} = 10{\log _{10}} \left( \mathbb{E} {\left[ {\left\| {{\bm{H}} - {\bm{\widehat H}}} \right\|_F^2/\left\| {\bm{H}} \right\|_F^2} \right]} \right), 
\end{align}
and the average spectral efficiency (ASE), defined as
\begin{align}\label{eq:ASE}
\begin{array}{rcl}
{\rm{ASE}} \! \! \! \! &=& \! \! \! \! {\log _2}{\rm{det}} \Big( {\bm{I}_{N_{\rm{RF}}}}\\ [2mm]
\! \! \! \! &\ & \! \! \! \! { + \!\  \dfrac{1}{{N_{\rm{RF}}}}{\bm{R}}_n^{-1}{\bm{W}}_{\rm{opt}}^H{\bm{H}}{{\bm{F}}_{\rm{opt}}}{{\bm{F}}_{\rm{opt}}^H}{{\bm{H}}^H}{{\bm{W}}_{\rm{opt}}}} \Big), 
\end{array}
\end{align}
where ${{\bm{R}}_n} \buildrel \Delta \over = \sigma _n^2{\bm{W}}_{\rm{opt}}^H{{\bm{W}}_{\rm{opt}}}$, and $\bm{F}_{\rm{opt}}$ and $\bm{W}_{\rm{opt}}$ are the optimal precoder and combiner consisting of the first $N_{\rm{RF}}$ columns of ${\bm{\widehat V}}$ and ${\bm{\widehat U}}$, respectively \cite{{OMP},{Subspace}}, and the given ${\bm{\widehat V}}$ and ${\bm{\widehat U}}$ are the left and right singular matrices of ${\bm{\widehat H}}$, i.e. ${\bm{\widehat H}} = {\bm{\widehat U\widehat \varSigma }}{{\bm{\widehat V}}^H}$.
Additionally, the bit error rate (BER) performance with the optimal precoder $\bm{F}_{\rm{opt}}$ and combiner $\bm{W}_{\rm{opt}}$ is also investigated, where the number of data streams used in downlink transmission is $N_{\rm{S}} = {N_{\rm{RF}}}$ and the modulation mode is 16-QAM.
Noted that we assume the MS has a full CSI estimated at the BS without considering the specific feedback mechanism here.

We begin by discussing the pilot overhead required among the proposed 2D unitary ESPRIT based channel estimation scheme, the ACS-based channel estimation scheme \cite{{ACS}}, and the OMP-based channel estimation scheme \cite{{OMP}}.
We consider the number of paths is $L = 5$ for example.
The corresponding pilot overhead are ${T_{\rm{Proposed}}} = {T_{\rm{MS}}}{N_{\rm{b}}^{\rm{R}}}{N_{\rm{b}}^{\rm{T}}} = 300$ for our proposed scheme, ${T_{\rm{ACS}}} = K{L^2}\left( {KL/N_{\rm{RF}}} \right){\log _K}\left( {{G_{\rm{ACS}}}/L} \right) = 1500$ with $K = 4$, $G_{\rm{ACS}} = 320$ for the ACS-based scheme, and ${T_{\rm{OMP}}} = N_{\rm{T}}^{\rm{Beam}}N_{\rm{R}}^{\rm{Beam}}/N_{\rm{RF}} = 576$ with $N_{\rm{T}}^{\rm{Beam}} = N_{\rm{R}}^{\rm{Beam}} = 48$ and $G_{\rm{OMP}} = 150$ for the OMP-based scheme, respectively.
Obviously, the pilot overhead required for our proposed scheme is the smallest.
Furthermore, the computational complexity of three different schemes will be compared with the specific simulation parameters.
According to the discussion in Section III-D and the simulation parameters used in Section IV, the computational complexity for the proposed scheme is ${C_{\rm{Proposed}}} = {\cal O}\left( {5850} \right)$.
While those for the ACS-based and the OMP-based channel estimation schemes are ${C_{\rm{ACS}}} = {\cal O}\left( {7.8 \times {{10}^6}} \right)$, and ${C_{\rm{OMP}}} =  {\cal O}\left( {5.28 \times {{10}^{7}}} \right)$ (here we consider ${\left| {{{\cal I}_t}} \right|} = 50$ in simulations), respectively.
Therefore, we have ${C_{\rm{Proposed}}}/{C_{\rm{ACS}}} = {7.5 \times 10^{-4}}$ and ${C_{\rm{Proposed}}}/{C_{\rm{OMP}}} = {1.1 \times 10^{-4}}$.
Clearly, the low complexity of our proposed scheme is self-evident.

\begin{figure}[!tp]
\centering
{\subfigure[]{ \label{fig:NMSE1(a)}
\centering
\begin{minipage}[b]{0.44\textwidth}
\includegraphics[width=1\textwidth]{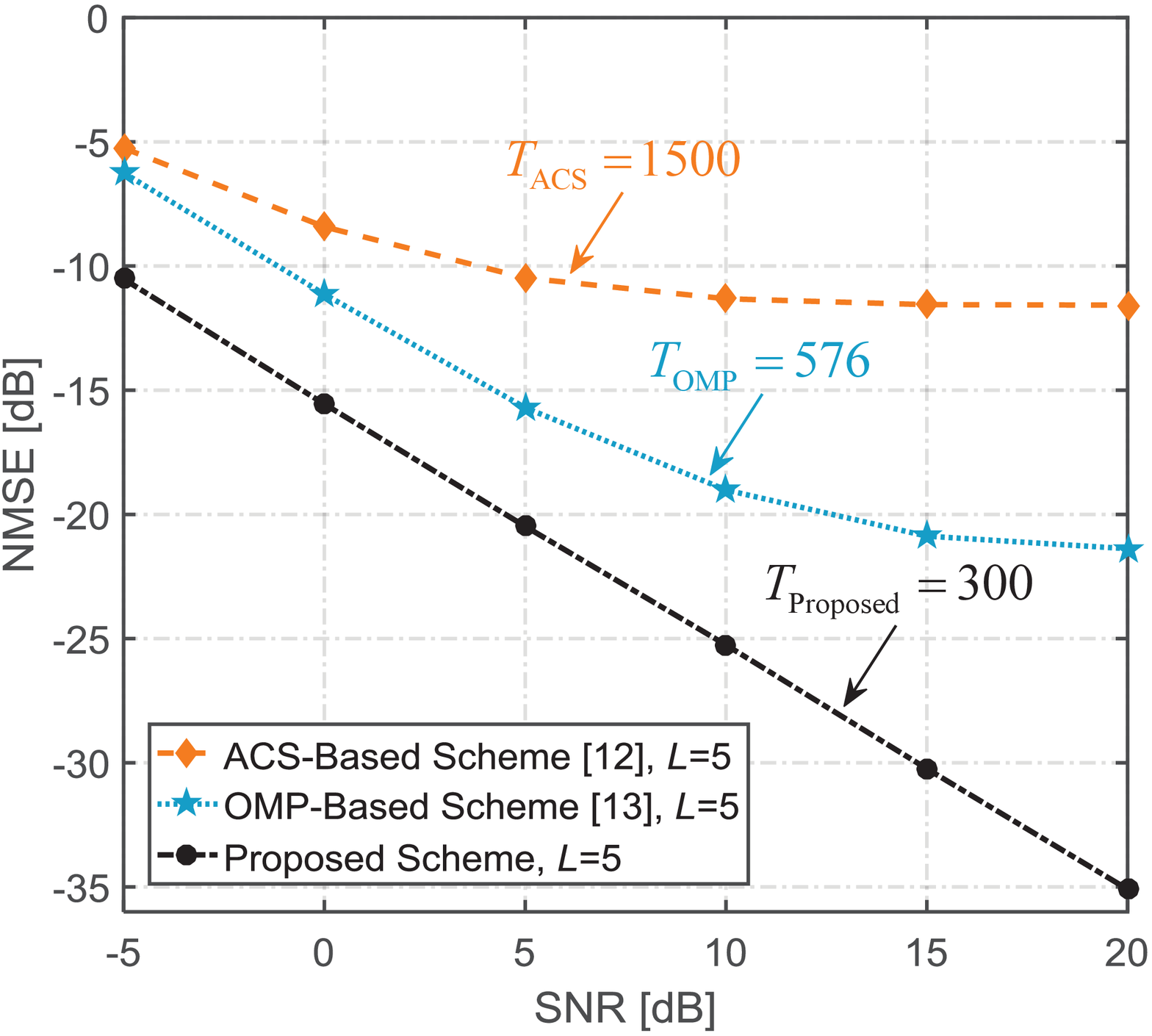}
\vspace{-4.5mm}
\end{minipage}
}}
{\subfigure[]{ \label{fig:NMSE1(b)}
\centering
\begin{minipage}[b]{0.44\textwidth}
\includegraphics[width=1\textwidth]{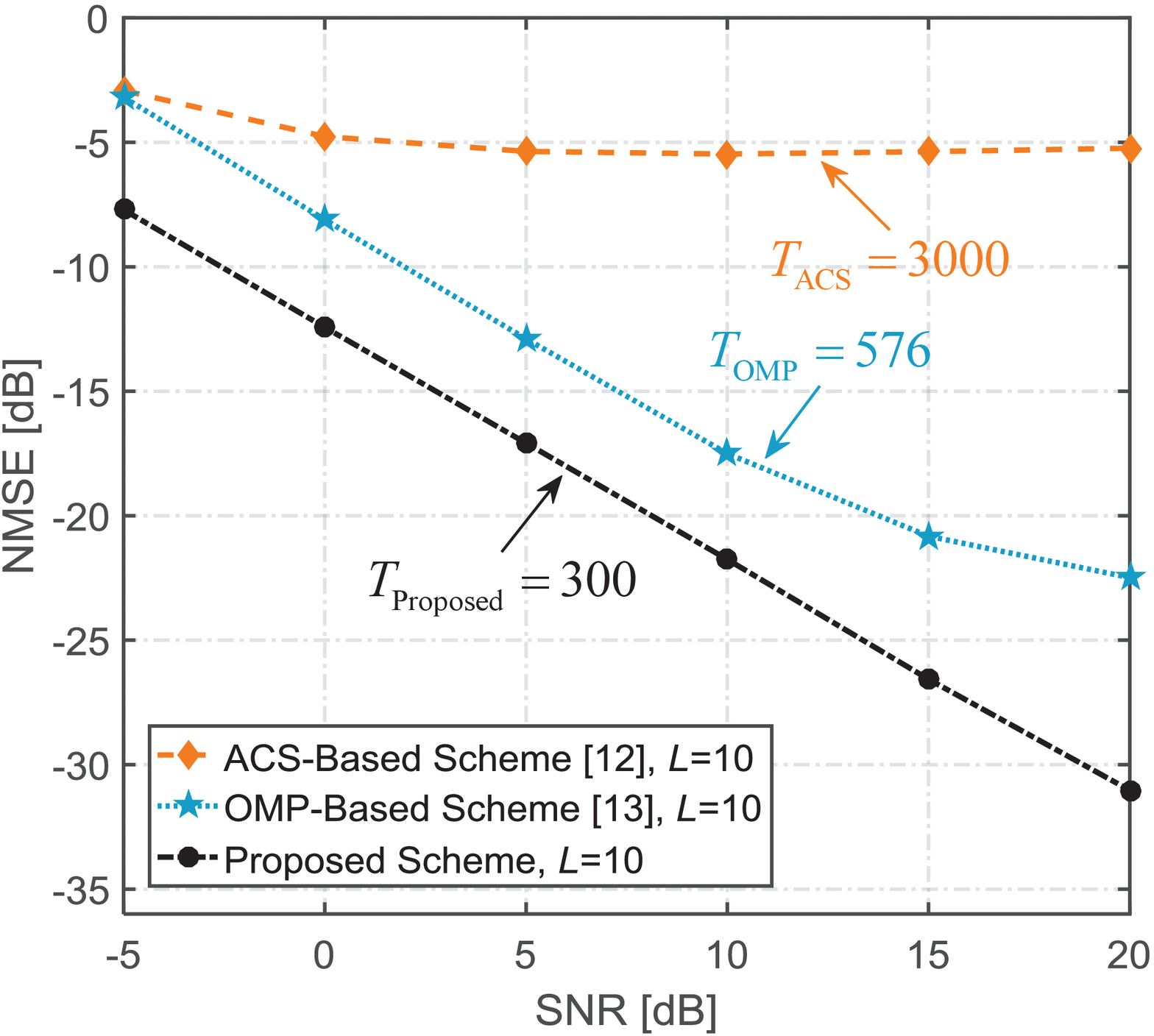}
\vspace{-4.5mm}
\end{minipage}
}}
\caption{NMSE performance comparison of different channel estimation schemes versus SNRs: (a) number of paths $L = 5$, (b) number of paths $L = 10$.} \label{fig:NMSE1}
\end{figure}

\begin{figure}[!tp]
     \centering
     \includegraphics[width=8.1cm, keepaspectratio]
     {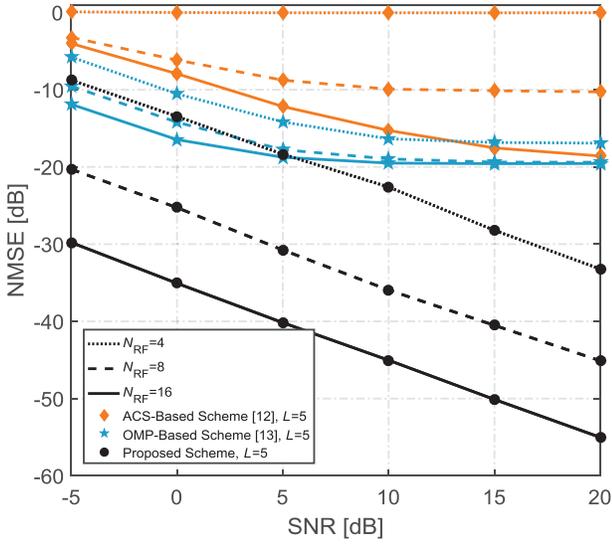}
     \caption{NMSE performance comparison of different schemes versus SNRs, where $N_{\rm{RF}} = 4, N_{\rm{RF}} = 8$ and $N_{\rm{RF}} = 16$ are considered, respectively.}
     \label{fig:NMSE2}
     \vspace*{-4.5mm}
\end{figure}

\begin{figure}[!tp]
     \centering
     \includegraphics[width=8.1cm, keepaspectratio]
     {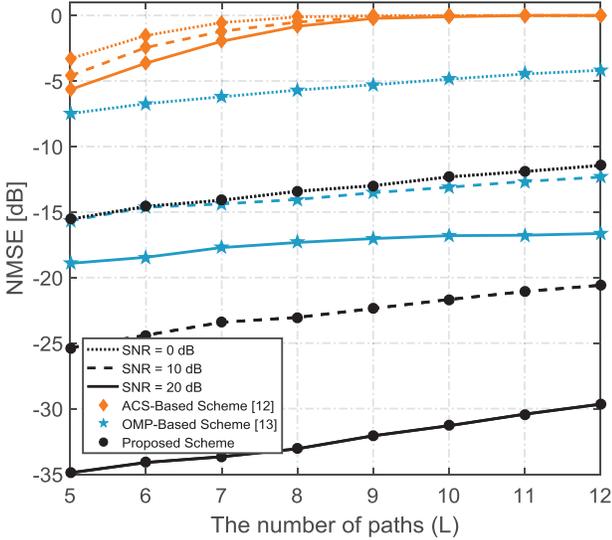}
     \caption{NMSE performance comparison of different schemes with the fixed pilot overhead against different numbers of paths, where SNRs are 0 dB, 10 dB, and 20 dB, respectively. }
     \label{fig:NMSE3}
     \vspace*{-4.5mm}
\end{figure}

\begin{figure}[!tp]
\centering
{\subfigure[]{ \label{fig:ASE(a)}
\begin{minipage}[b]{0.44\textwidth}
\includegraphics[width=1\textwidth]{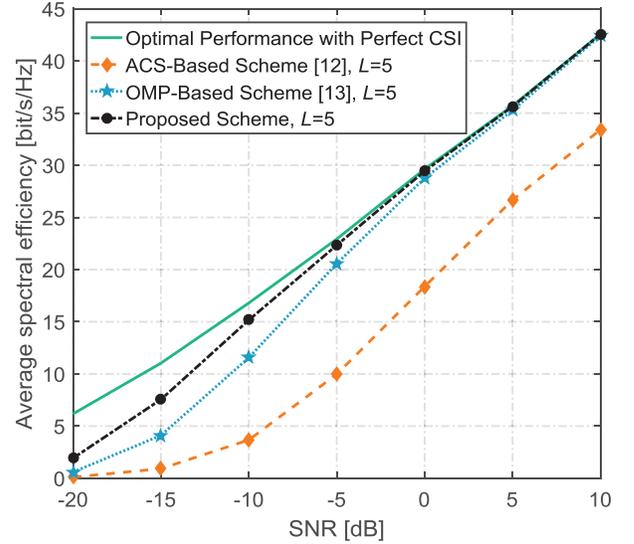}
\vspace{-4.5mm}
\end{minipage}
}}
{\subfigure[]{ \label{fig:ASE(b)}
\begin{minipage}[b]{0.44\textwidth}
\includegraphics[width=1\textwidth]{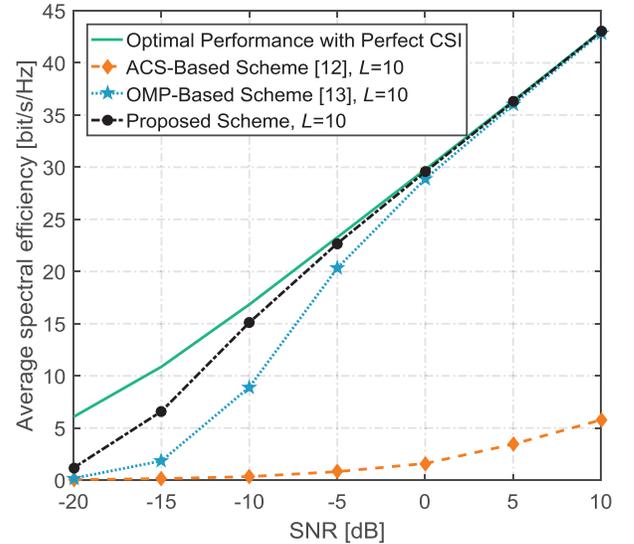}
\vspace{-4.5mm}
\end{minipage}
}}
\caption{Comparison of ASE performance among different channel estimation schemes versus SNRs, where the optimal performance with the perfect CSI known at both BS and MS are considered as the upper bound: (a) number of paths $L = 5$, (b) number of paths $L = 10$.}\label{fig:ASE}
\end{figure}

\begin{figure}[!tp]
\centering
{\subfigure[]{ \label{fig:BER(a)}
\begin{minipage}[b]{0.44\textwidth}
\includegraphics[width=1\textwidth]{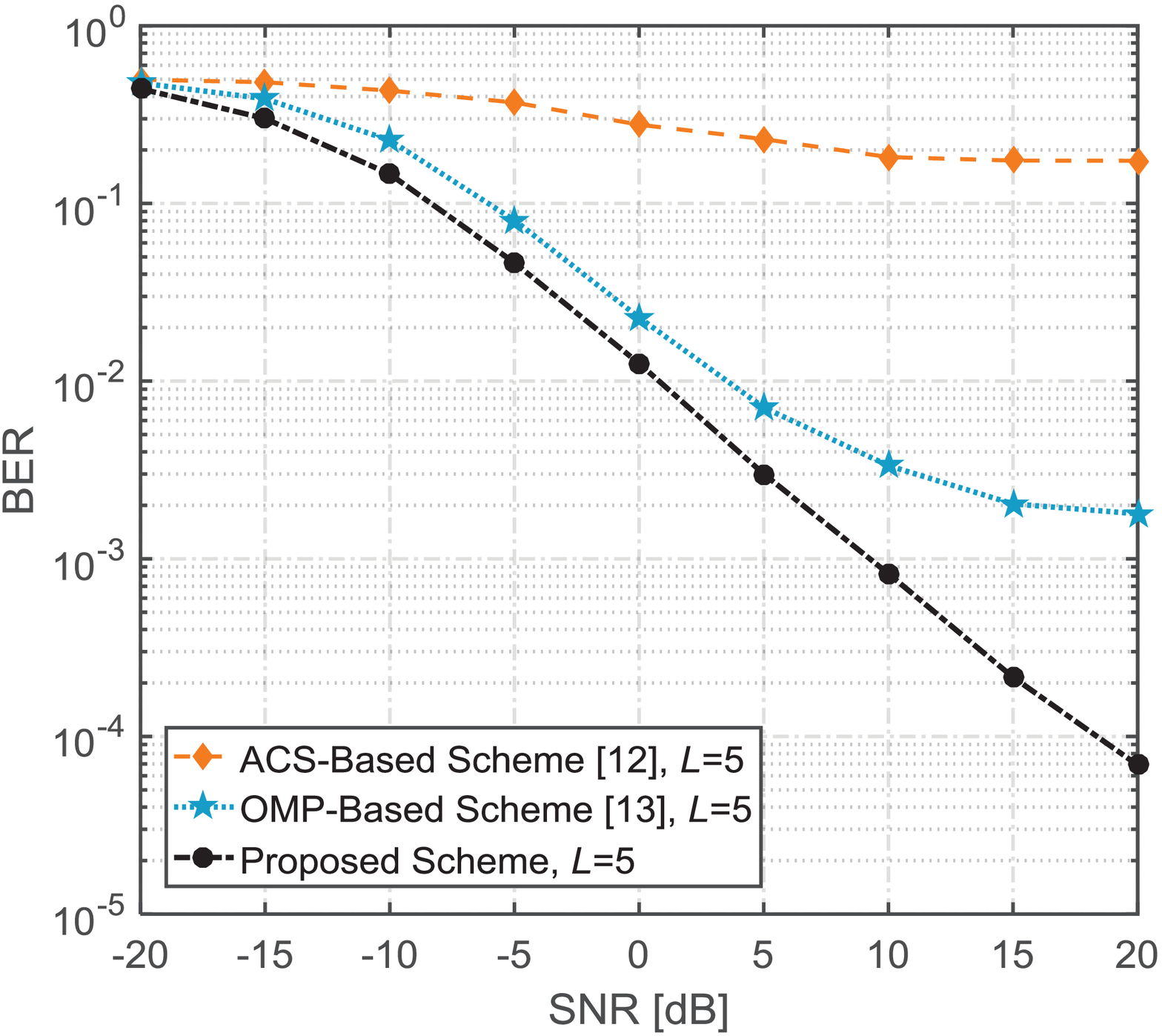}
\vspace{-4.5mm}
\end{minipage}
}}
{\subfigure[]{ \label{fig:BER(b)}
\begin{minipage}[b]{0.44\textwidth}
\includegraphics[width=1\textwidth]{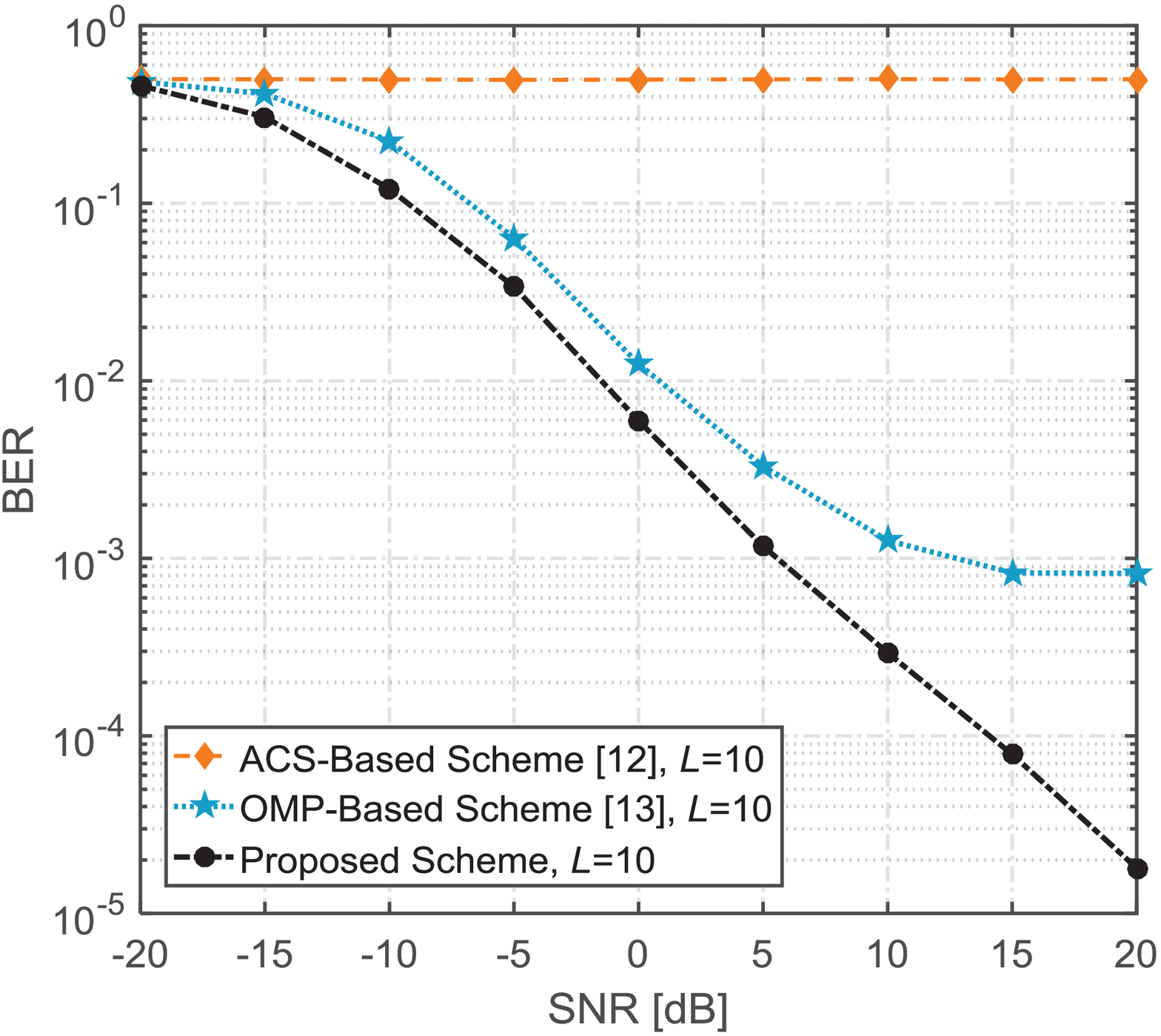}
\vspace{-4.5mm}
\end{minipage}
}}
\caption{BER performance comparison of different schemes versus SNRs, where $N_{\rm{S}} = {N_{\rm{RF}}}$, 16-QAM, and the optimal precoder and combiner are considered: (a) number of paths $L = 5$, (b) number of paths $L = 10$.}\label{fig:BER}
\end{figure}

Fig. \ref{fig:NMSE1(a)} and \ref{fig:NMSE1(b)} compare the NMSE performance of the proposed 2D unitary ESPRIT based channel estimation scheme with the ACS-based and the OMP-based channel estimation schemes against different signal-to-noise ratios (SNRs) with $L = 5$ and $L = 10$, respectively.
Additionally, the pilot overhead required for different schemes is also provided as ${T_{\rm{ACS}}} = 1500$ (with $L = 5$) and ${T_{\rm{ACS}}} = 3000$ (with $L = 10$) for the ACS-based scheme, ${T_{\rm{OMP}}} = 576$ for the OMP-based scheme, and ${T_{\rm{Proposed}}} = 300$ for the proposed scheme.
From Fig. \ref{fig:NMSE1(a)}, it can be observed that the NMSE performance of our proposed scheme outperforms the other two schemes significantly with a much reduced pilot overhead.
Moreover, the performance gap between the proposed scheme and its counterparts becomes larger when SNR increases.
Especially, the NMSE performance of the proposed scheme are more than 10 dB and 5 dB better than the ACS-based and OMP-based schemes, respectively.
This is because the proposed channel estimation scheme can acquire the super-resolution estimates of the AoAs and AoDs with high accuracy.
By contrast, the ACS-based and the OMP-based channel estimation schemes suffer from the obvious performance floor when SNR becomes large.
This is because for the ACS-based channel estimation scheme, the estimation resolution of the AoAs and AoDs is limited by the size of codebook and the resolution of quantized grids.
It should be pointed out that the ACS-based channel estimation scheme can not effectively distinguish multiple AoAs or AoDs close to each other, and will work poorly when the number of paths becomes large\footnote{In simulations, we use the MATLAB codes provided by the authors in http://www.aalkhateeb.net/publications.html?i=1}.
From Fig. \ref{fig:NMSE1(b)}, we can observe that when $L=10$ at SNR = 20 dB, the NMSE performance of the ACS-based channel estimation scheme is just around -5 dB.
While for the OMP-based channel estimation scheme quantizing the continuously distributed AoAs and AoDs as the discretized grids, the NMSE performance will have the floor effect at high SNR due to the limited estimate resolution of the AoAs and AoDs.
It is also worth pointing out again that our proposed scheme requires a much reduced pilot overhead compared with two other channel estimation schemes.
This means that, to achieve the better NMSE performance, the proposed scheme can reduce the required pilot overhead by $80\%$ and $48\%$, respectively, compared with the ACS-based and the OMP-based channel estimation schemes.

Fig. \ref{fig:NMSE2}. compares the NMSE performance of different channel estimation schemes versus SNRs, where ${T_{\rm{ACS}}} = 312$, ${T_{\rm{OMP}}} = 256$, ${T_{\rm{Proposed}}} = 243$, $N_{\rm{RF}} = 4, N_{\rm{RF}} = 8$, $N_{\rm{RF}} = 16$, and $L = 5$ are considered.
From Fig. \ref{fig:NMSE2}, we can observe that the NMSE performance of the proposed scheme improve considerably when the number of RF chains increases from $N_{\rm{RF}} = 4$ to $N_{\rm{RF}} = 16$.
This is because the larger number of RF chains $N_{\rm{RF}}$ can enlarge the effective observation dimension and thus improve the NMSE performance.
By contrast, when $N_{\rm{RF}}$ increases, the NMSE performance of both the ACS-based and the OMP-based channel estimation schemes improves slightly due to the floor effect.
Particularly, if SNR = 10 dB is considered, it can be observed that for the proposed scheme, the ACS-based scheme, and the OMP-based scheme, the performance improvements are 22 dB, 15 dB, and 5 dB, respectively, when $N_{\rm{RF}}$ increases from 4 to 16.
It is worth pointing out that the NMSE performance improvement of the OMP-based channel estimation scheme can be negligible as $N_{\rm{RF}}$ becomes large, since the accuracy of channel estimation heavily depends on the resolution of the quantized grids $G_{\rm{OMP}}$ rather than the number of RF chains $N_{\rm{RF}}$.
This phenomenon further confirms the fact that quantizing the continuously distributed AoAs and AoDs for channel estimation will lead to an inevitable quantization error and thus a non-negligible performance loss.

Fig. \ref{fig:NMSE3}. compares the NMSE performance of different schemes with the fixed pilot overhead against different numbers of paths, where ${T_{\rm{ACS}}} = 375$, ${T_{\rm{OMP}}} = 256$, ${T_{\rm{Proposed}}} = 243$.
It shows that the NMSE performance of all schemes decreases when the number of paths increases.
Fig. \ref{fig:NMSE3} further confirms the better performance of the proposed scheme than the ACS-based and OMP-based schemes for different numbers of paths with different SNRs.

Fig. \ref{fig:ASE}. compares the ASE performance of different channel estimation schemes against different SNRs, where ${T_{\rm{ACS}}} = 375$, ${T_{\rm{OMP}}} = 256$, ${T_{\rm{Proposed}}} = 243$, and both $L=5$ and $L=10$ are considered.
In Fig. \ref{fig:ASE}, the ASE with the perfect CSI known at BS and MS is considered as the performance upper bound.
It can be observed from Fig. \ref{fig:ASE(a)} that our proposed scheme is superior to the ACS-based and OMP-based channel estimation schemes, and its performance approaches the optimal performance when SNR is larger than -5 dB.
The ASE performance of the ACS-based channel estimation scheme is rather poor for $L = 10$, since it can not effectively estimate the AoAs and AoDs when the number of paths becomes large.
It is worthy pointing out that although the performance gap between the proposed scheme and the OMP-based scheme becomes smaller as SNR increases, the computational complexity of the proposed scheme is much smaller than that of the OMP-based scheme.

Fig. \ref{fig:BER}. compares the BER performance of three different schemes versus SNRs.
In Fig. \ref{fig:BER}, we consider ${T_{\rm{ACS}}} = 375$, ${T_{\rm{OMP}}} = 256$, ${T_{\rm{Proposed}}} = 243$, the number of data streams used in downlink transmission is $N_{\rm{S}} = {N_{\rm{RF}}}$, the modulation type is 16-QAM, and the optimal precoder and combiner are adopted by the BS and MS.
Obviously, the BER performance of our proposed scheme is better than the ACS-based and OMP-based scheme.
The OMP-based scheme suffers from an obvious BER performance floor when SNR is larger than 10 dB for both $L=5$ and $L=10$, while the ACS-based scheme cannot work when $L=10$.
Additionally, it is worthy pointing out that although the larger number of paths will lead to the worse NMSE performance for both the proposed scheme and the OMP-based scheme, their BER performance  improves when $L$ increases from 5 to 10.
This is because the larger number of paths can provide the higher spatial diversity gains, and thus can improve the BER performance.

\section{Conclusions}
In this paper, we have proposed a 2D unitary ESPRIT based super-resolution channel estimation scheme for the mmWave massive MIMO systems with hybrid precoding.
The proposed scheme can jointly obtain the super-resolution estimates of AoAs and AoDs with high accuracy.
Specifically, by designing an efficient uplink training signals at both BS and MS, we can use a much reduced pilot overhead to acquire the low-dimensional effective channel, which has the same shift-invariance of array response as the high-dimensional mmWave MIMO channel to be estimated.
Then, by exploiting the 2D unitary ESPRIT based channel estimation algorithm, the super-resolution estimates of AoAs and AoDs can be jointly acquired from the low-dimensional effective channel.
Moreover, the associated path gains can be acquired by using the LS estimator.
Finally, the high-dimensional mmWave MIMO channel can be reconstructed according to the obtained AoDs, AoDs, and path gains.
Simulation results have verified that the proposed channel estimation scheme can achieve the better channel estimation performance with lower pilot overhead and computational complexity than the conventional schemes.


\begin{thebibliography}{18}

\bibitem{overview:Heath}
R. W. Heath, Jr., N. Gonz\'{a}lez-Prelcic, S. Rangan, W. Roh, and A. M. Sayeed,
``An overview of signal processing techniques for millimeter wave MIMO systems,"
{\em IEEE J. Sel. Topics Signal Process.}, vol. 10, no. 3, pp. 436-453, Apr. 2016.

\bibitem{Energy:efficientGao}
X. Gao, L. Dai, S. Han, C.-L. I, and R. Heath,
``Energy-efficient hybrid analog and digital precoding for mmWave MIMO systems with large antenna arrays,"
{\em IEEE J. Sel. Areas Commun.}, vol. 34, no. 4, pp. 998-1009, Apr. 2016.

\bibitem{MIMO:PCS}
A. Alkhateeb, J. Mo, N. G. Prelcic, and R. W. Heath Jr.,
``MIMO precoding and combining solutions for millimeter-wave systems,"
{\em IEEE Commun. Mag.}, vol. 52, no. 12, pp. 122-131, Dec. 2014.

\bibitem{Hybrid:MIMO}
R. M. Rial, C. Rusu, N. G. Prelcic, A. Alkhateeb, and R. W. Heath Jr.,
``Hybrid MIMO architectures for millimeter wave communications: Phase shifters or switches?,"
{\em IEEE Access}, vol. 4, pp. 247-267, Jan. 2016.


\bibitem{LSE}
C. Huang, L. Liu, C. Yuen, and S. Sun,
``A LSE and sparse message passing-based channel estimation for mmWave MIMO systems,"
in {\em Proc. IEEE Globecom Workshops (GC Wkshps)}, Dec. 2016, pp. 1-6.

\bibitem{Overlapped}
M. Kokshoorn, H. Chen, P. Wang, Y. Li, B. Vucetic,
``Millimeter wave MIMO channel estimation using overlapped beam patterns and rate adaptation,"
{\em IEEE Trans. Signal Process.}, vol. 65, no. 3, pp. 601-616, Feb. 2017.



\bibitem{Beamspace}
L. Dai, X. Gao, S. Han, C.-L. I, and X. Wang,
``Beamspace channel estimation for millimeter-wave massive MIMO systems with lens antenna array,"
in {\em Proc. IEEE/CIC Int. Conf. Commun. China (ICCC)}, Jul. 2016, pp. 1-6.

\bibitem{LimitedRF}
L. Yang, Y. Zeng, R. Zhang,
``Effcient channel estimation for millimeter wave MIMO with limited RF chains,"
in {\em Proc. IEEE Int. Conf. Commun. (ICC)}, May 2016, pp. 1-6.


\bibitem{Auxiliary}
D. Zhu, J. Choi, and R. W. Heath, Jr.,
``Auxiliary beam pair enabled AoD and AoA estimation in closed-loop large-scale millimeter-wave MIMO systems,"
{\em IEEE Trans. Wireless Commun.}, vol. 16, no. 7, pp. 4770-4785, Jul. 2017.

\bibitem{Subspace}
H. Ghauch, T. Kim, M. Bengtsson, and M. Skoglund,
``Subspace estimation and decomposition for large millimeter-wave MIMO systems,"
{\em IEEE J. Sel. Top. Signal Process.}, vol. 10, no. 3, pp. 528-542, Apr. 2016.


\bibitem{AdaptiveCS}
S. Sun and T. S. Rappaport,
``Millimeter wave MIMO channel estimation based on adaptive compressed sensing,"
in {\em Proc. IEEE Int. Conf. Commun. Workshops (ICC Workshops)}, May 2017, pp. 47-53.

\bibitem{ACS}
A. Alkhateeb, O. El Ayach, G. Leus, and R. W. Heath, Jr.,
``Channel estimation and hybrid precoding for millimeter wave cellular systems,"
{\em IEEE J. Sel. Topics Signal Process.}, vol. 8, no. 5, pp. 831-846, Oct. 2014.

\bibitem{OMP}
J. Lee, G. T. Gil, and Y. H. Lee,
``Channel estimation via orthogonal matching pursuit for hybrid MIMO systems in millimeter wave communications,"
{\em IEEE Trans. Commun.}, vol. 64, no. 6, pp. 2370-2386, Jun. 2016.


\bibitem{esprit}
R. Roy and T. Kailath,
``ESPRIT-estimation of signal parameters via rotational invariance techniques,"
{\em IEEE Trans. Acoust., Speech, Signal Process.}, vol. 37, no. 7, pp. 984-995, Jul. 1989.

\bibitem{2Du:esprit}
M. D. Zoltowski, M. Haardt, and C. P. Mathews,
``Closed-form 2-D angle estimation with rectangular arrays in element space or beamspace via unitary ESPRIT,"
{\em IEEE Trans. Signal Process.}, vol. 44, no. 2, pp. 316-328, Feb. 1996.

\bibitem{JADE:SIT}
A. van der Veen, M. Vanderveen, and A. Paulraj,
``Joint angle and delay estimation using shift-invariance techniques,"
{\em IEEE Trans. Signal Process.}, vol. 46, no. 2, pp. 405-418, Feb. 1998.

\bibitem{JADE}
H. Miao, M. Juntti, and K. Yu,
``2-D unitary ESPRIT based joint AOA and AOD estimation for MIMO system,"
in {\em Proc. IEEE 17th Int. Symp. PIMRC}, Sep. 2006, pp. 1-5.


\bibitem{Matrix}
G. H. Golub and C. F. Van Loan,
\emph{Matrix Computations (4rd ed.).}
Baltimore, MD, USA: The Johns Hopkins Univ. Press, 2013.


\end{thebibliography}
\end{document}